\documentclass[
reprint,
superscriptaddress,
aps,
pre,
]{revtex4-2}

\usepackage{ifthen}
\usepackage{amsmath}
\usepackage{amssymb}
\usepackage{makeidx}
\usepackage[colorlinks=true,citecolor=blue,linkcolor=blue,linktocpage=true,pagebackref=false]{hyperref}
\hypersetup{colorlinks=true,citecolor=blue,linkcolor=blue,filecolor=blue,urlcolor=blue}
\usepackage{soul}
\usepackage{color}
\usepackage{graphicx}
\usepackage{dcolumn}
\usepackage[T1]{fontenc}
\usepackage{wasysym}

\usepackage{bm}
\usepackage{physics}
\usepackage[normalem]{ulem}

\begin{document}

\title{
Ensemble-averaged mean-field many-body level density: \\ an indicator of integrable versus chaotic single-particle dynamics
}
\author{Georg Maier}
\affiliation{Institut f\"ur Theoretische Physik, Universit\"at Regensburg, D-93040 Regensburg, Germany}

\author{Carolyn Echter}
\affiliation{Institut f\"ur Theoretische Physik, Universit\"at Regensburg, D-93040 Regensburg, Germany}

\author{Juan Diego Urbina}
\affiliation{Institut f\"ur Theoretische Physik, Universit\"at Regensburg, D-93040 Regensburg, Germany}

\author{Caio Lewenkopf}
\affiliation{Instituto de F\'{\i}sica, Universidade Federal do Rio de Janeiro, 
             21941-972 Rio de Janeiro, RJ, Brazil}
             
\author{Klaus Richter}
\affiliation{Institut f\"ur Theoretische Physik, Universit\"at Regensburg, D-93040 Regensburg, Germany}

\date{\today}

\begin{abstract}
According to the quantum chaos paradigm, the nature of a system's classical dynamics, whether integrable or chaotic, is universally reflected in the {\it fluctuations} of its quantum spectrum.
However, since many-body spectra in the mean field limit are composed of independent single-particle energy levels,
their spectral fluctuations always display Poissonian behavior and hence cannot be used to distinguish underlying chaotic from integrable single-particle dynamics. 
We demonstrate that this distinction can, instead, be revealed from the {\it mean} many-body level density (at fixed energy) and its variance after averaging over ensembles representing different types of single-particle dynamics. 
This is in strong contrast to the energy-averaged {\it mean} level density (of a given system)
that is assumed not to carry such information and is routinely removed to focus on universal signatures. 
To support our claim we systematically analyze the role of 
single-particle level correlations, that enter through Poisson and random matrix statistics (of various symmetry classes)
into the ensemble-averaged density of states and its variance, contrasting
bosonic and fermionic many-body systems.
Our analytical study, together with extensive numerical simulations for systems with $N \ge 5$ particles consistently reveal significant differences 
(up to an order of magnitude
for fermions and even larger for bosons) in the mean many-body level densities, depending on the nature of the underlying dynamics. 
Notably,  in the fermionic case Poisson-type single-particle level fluctuations precisely cancel contributions from indistinguishability, such that the average many-body spectral density equals the (Thomas-Fermi) volume term.
We further highlight the difference between the mean level density and its variance as functions of the total energy $E$ and the excitation energy $Q$.
\end{abstract}

\maketitle

\section{Introduction}
\label{sec:introduction}

The ideas and methods that gave rise to what is now known as the field of quantum chaos \cite{gutzwiller2013,Stöckmann_1999}, developed during the 70-90's and reaching its modern form during the 2000's \cite{haake2001,richter2022}, have provided a comprehensive understanding of the way classical phase space structures manifest in the quantum properties of physical systems \cite{Gutzwiller:2007_schoolarpedia}.
A cornerstone of this field is the Bohigas-Giannoni-Schmit (BGS) conjecture \cite{BGS1984}, which asserts that for the two extreme types of classical dynamics—precisely defined as the fully integrable and chaotic regimes—the statistical fluctuations in the high-lying regions of the spectrum are universal, i.e., independent of the specific system. 
These fluctuations follow either Poisson statistics or correspond to those of the ensembles in Random Matrix Theory (RMT) \cite{mehta2004random, Livan2018, Potters2020}, depending on the appropriate universality class \cite{Dyson1962, Altland1997}.

Although the BGS conjecture was originally formulated for Single-Particle (SP) systems, this concept of quantum chaos is clearly applicable to Many-Body (MB) systems as well, as long as they have a classical limit. The well-defined notions of chaos and integrability remain fully independent of the dimension of the classical phase space, and therefore, of the (finite) number of degrees of freedom \cite{Almeida_1989,Tabor1989ChaosAI}.

Nevertheless, the emergence of new time and energy scales \cite{Schiulaz} makes the connection with RMT/Poissonian ensembles more subtle in the MB case \cite{Kos_2018,Chan_2018}. 
The study of the interplay between the traditional signatures of quantum chaos and these new scales, mediated by intrinsic MB properties like locality of physical interactions, and even in systems without a semiclassical regime, is the subject of MB quantum chaos.

The idea that the specific way particles interact requires a modification of the connection between chaos and RMT was already evident in the 80's. 
By that time, special ensembles designed to correctly account for the two-body nature of interacting potentials were introduced under the name of embedded random matrix ensembles \cite{book}, and found important applications in nuclear physics \cite{Papenbrock2007}. 
With the advent of MB chaos, the construction of generalized RMT ensembles became a standard practice, playing a significant role, across a wide range of fields from MB localization \cite{Buijsman_2019} to quantum circuits \cite{Fisher_2023}.

An important limit of an ensemble of random matrices with a special structure occurs when the interactions are described instead within a mean-field approach, namely, when the MB system is approximated by a 
gas of quasiparticles. 
In this situation a suitable ensemble is constructed where the MB Hamiltonian is devised as the direct sum of one-body operators, one for each independent effective degree of freedom.
The dynamical features of the SP dynamics (their integrable  or chaotic behavior) are then incorporated by the choice of the ensemble from which these SP Hamiltonians are selected, namely the Poisson and Gaussian ensembles for integrable and chaotic dynamics, respectively. 

In general, in MB systems, there is wide range of possible combinations where at the mean-field 
level, a subset of degrees of freedom exhibit integrable dynamics while another subset is chaotic \cite{Hämmerling2010, Hämmerling2011, Freese2016}. 
A major simplification occurs, however, when we focus on systems of identical (quasi) particles. 
MB systems of non-interacting but SP-chaotic and identical particles must be described by a single SP Hamiltonian within the universality class that defines the SP RMT/Poisson ensemble.
In the case of finite-dimensional SP Hilbert spaces, such as lattice systems with non-interacting (quadratic) Hamiltonians, we obtain the non-interacting limit of Bose and Fermi-Hubbard models with RMT/Poisson SP spectra, a type of systems that has received much attention recently \cite{Liao_2020,łydżba2024,lev2021}.

Specifically, for this scenario of  non-interacting 
or mean-field systems (of indistinguishable particles) with individually chaotic or integrable degrees of freedom, the construction of an appropriate statistical approach is formulated in two steps. 
First, the MB physical property of interest is expressed in terms of an arbitrary but fixed SP Hamiltonian $\hat{H}_{\rm sp}$. 
This is possible because, for 
mean-field
systems, all MB observables can be expressed in terms of SP objects.
In the case of interest in this paper, the MB  density of states, for example, can be expressed using combinatorial methods that start from the eigenvalues of $\hat{H}_{\rm sp}$.
Second, the nature of classical SP dynamics, in the extremes of SP integrability and chaos, is incorporated through the choice of the ensemble of SP Hamiltonians. 
Specifically this involves the so-called Poisson ensemble of uncorrelated SP energies for the integrable case, and one of the Gaussian ensembles of RMT for the chaotic counterpart  \cite{gutzwiller2013, Stöckmann_1999, haake2001}. 

When focusing on the spectral properties of this type of systems, one encounters an interesting situation from the perspective of the usual application of quantum chaos concepts.
Historically, focus has been on the signatures of integrability and chaos in the \textit{fluctuations} of the spectral density around its mean value.
However, by its very definition, the MB spectra of non-interacting systems or mean-field models of indistinguishable particles display Poissonian spectral fluctuations, due to the explicit integrability of its mean-field (classical) limit, even if the SP dynamics is chaotic.
As a consequence, the usual indicators of integrability and chaos are simply insensitive to the particular type of classical dynamics at the SP level. 

Still, the MB spectrum displays, as we demonstrate here, a strong sensitivity on the type of SP spectral fluctuations, which in turn, reflect the nature of the SP classical phase space.
The reason for this seemingly paradoxical situation lies, once again, in the unique properties of the ensemble of Hamiltonian matrices that are suitable for describing systems of independent
particles.
Unlike their SP counterparts (the RMT and Poisson ensembles that describe systems with chaotic or integrable classical limits) these ensembles do not exhibit a typical RMT ergodic behavior.

More precisely, this lack of ergodicity implies the non-equivalence between running averages (over small energy windows) of a given MB system on the one hand and averages over ensembles of systems (at fixed energy) on the other hand. The latter ensemble average is essential for comparing the predictions of RMT with the generic spectral properties of specific systems.
In other words: While ergodicity on the SP level is not reflected in the fluctuations of the mean-field MB level density, there is a strong dependence of the mean level density (in the sense of ensemble averages) on the SP ensemble, to the extent that it behaves completely differently from the smoothed level density obtained by local convolution within some energy window.

For illustration consider as a prominent example mean-field fermions in a closed disordered mesoscopic quantum system. In the localized and delocalized (conducting) regime, the SP spectra obey Poisson and RMT statistics, respectively. A disorder average (at fixed energy), commonly performed in mesoscopic physics, represents the ensemble average, in contrast to the energy or temperature smoothing of an individual system with given disorder configuration.

In this paper, we present an in-depth study of such an ensemble average and demonstrate how the quantum signatures of chaos or integrability in the SP dynamics impact the ensemble-averaged MB density of states (MBDOS) and its variance for
systems of identical particles, e.g., fermions and bosons in a mean-field framework.  
We analyze the MBDOS in two ways:
First, we study the ensemble-averaged MBDOS as a function of the system's total energy $E$, a setting that is more amenable for an analytical approach. 
Next, we analyze the MBDOS as a function of the excitation energy, $Q=E-E_{\rm GS}$, where $E_{\rm GS}$
is the ground-state energy.
For this construction, which is closer to experimental information, we calculate the mean level density from an ensemble of systems, subtracting $E_{\rm GS}$ from each realization before averaging.
We demonstrate that eliminating the realization-to-realization dependence on $E_{\rm GS}$ significantly impacts the results.

Our findings are based on a combination of analytical results and extensive numerical simulations.

The paper is organized as follows. In Sec.~\ref{sec:theory} 
we review the Weidenm\"uller convolution formula \cite{Weidenmuller1993} for the MBDOS and present both the analytical and the numerical approaches we use to obtain the ensemble-average MBDOS.
In Secs.~\ref{sec:results_integrable} and ~\ref{sec:Chaos} we calculate the average MBDOS  for Poisson statistics and 
for the random matrix universal symmetry classes, namely, the orthogonal $(\beta = 1)$, unitary $(\beta = 2)$, and symplectic $(\beta = 4)$ Gaussian ensembles,
and consider in particular the dependence on energy $E$ and number of particles $N$. 
In Sec.~\ref{sec:signatures} we analyze the cumulative mean MBDOS as a function of the excitation energy $Q$ and discuss its dependence on the nature of the underlying SP dynamics.
We conclude in Sec.~\ref{sec:conclusions}.

\section{Theory and methods}
\label{sec:theory}

The study of the MBDOS of a Fermi gas has a very long history dating back to the pioneering work of Bethe \cite{Bethe1936}. 
The standard method \cite{Bethe1936,BohrMottelson1998} to compute the MBDOS is based on the  Laplace transform of the partition function and its evaluation using the Sommerfeld integral. 
More recently, it has been shown that the last step is not necessary, making it possible to calculate the MBDOS of boson gases \cite{Leboeuf2005}. 
Several improvements have been reported over the years \cite{Lefevre2023}. Unfortunately, the standard method does not allow for the inclusion SP spectral fluctuations. 

For this reason we use a different starting point in our analytical approach, namely, the Weidenmüller convolution formula for the MBDOS  \cite{Weidenmuller1993, Sommermann_1993}. 
We then write down its ensemble average, which depends on the corresponding SP fluctuations, and finally describe the numerical procedure used.

\subsection{MBDOS expressed in terms of SPDOS using the Weidenmüller convolution formula}

The MBDOS  $\rho(N, E)$ of systems with $N$ indistinguishable particles and MB energy $E$ can be obtained by calculating the Laplace transform of the (anti-)symmetrized MB propagator \cite{Weidenmuller1993}. 
For non-interacting particles, the MB propagator factorizes and the MBDOS decomposes into convolutions of the SPDOS $\rho(\varepsilon)$. For clarity, we denote  the SP energies by $\varepsilon$. 

When $N_i$ particles share the same energy, the (an\-ti-) symmetrization must be taken into account. The resulting MBDOS can be structured as a sum over all possible cycle decompositions of the symmetric group $S_N$, where all $N_i$ particles within a cycle share the same energy $E_i/N_i$, namely

\begin{widetext}
\begin{equation}
\label{convolution}
        \rho^{(\pm)}(N,E)=\frac{1}{N!} \sum_{l=1}^N (\pm 1)^{N-l} \!\!\!\sum_{\substack{N_1,...,N_l = 1 \\ N_1 \leq ... \leq N_l \\ \sum_{i=1}^l N_i = N}}^N c(N_1,...,N_l) 
        \prod_{i=1}^l \frac{1}{N_i} \int_0^\infty \text{d}E_1 \cdots \int_0^\infty \text{d}E_l~\delta\!\left(E - \sum_{i=1}^l E_i\right) \prod_{i=1}^l \rho\!\left(\frac{E_i}{N_i}\right).
\end{equation}
\end{widetext}
Here, $+\,(-)$ stands for bosonic (fermionic) systems, and $c(N_1,\dots,N_l)$ is the number of permutations with cycle decomposition consisting of $l$ cycles of lengths $N_1 , ..., N_l$, 
as given by \cite{Sommermann_1993}
\begin{equation}\label{permutations_in_cycle}
    c(N_1,\dots,N_l)=\frac{N!}{\prod\limits_{N_i \in \left\{N_1,...,N_l\right\}} m\!\left(N_i\right)! \prod\limits_{i=1}^l N_i},
\end{equation}
with $m(N_i)$ being the multiplicity of the cycle length $N_i$ within the cycle decomposition. 
These expressions are the key elements necessary for our study.
A detailed derivation of Eq.~\eqref{convolution} can be found in Refs.~\cite{Weidenmuller1993,Quirin}.

\subsection{Statistical approach to the average MBDOS}

We now compute the ensemble-averaged non-interacting MBDOS $\langle\rho^{(\pm)}(N,E)\rangle$ starting from Eq.~\eqref{convolution}.
The ensemble comprises different sets of eigenvalues $e_1,\dots, e_M$ of $M \times M$ matrices drawn form the corresponding matrix space. 
In this work, we consider Gaussian matrix ensembles and Poisson-distributed eigenvalues.

We denote the joint probability density of eigenvalues as $P_M(e_1,...,e_M)$ \cite{mehta2004random, bohigas1991random}. 
To address systems with a mean SP energy spacing $\rho_0^{-1}$, we need to transform the semi-circular DOS of the random matrix eigenvalue spectra $\{e_i\}$ into a constant DOS one. 
This is achieved by a process called unfolding \cite{bohigas1991random}. In practice, for large $M$, one can assume that random matrix spectra exhibit a constant mean level spacing $\bar{s}_M$ near the center of the band. 
Consequently, one can relate the energy levels $\varepsilon_i$ of a SP system with the eigenvalues $e_i$ of a random matrix by
\begin{equation}\label{e_to_eps}
    e_i/\bar{s}_M  = \varepsilon_i\rho_0.   
\end{equation}
As a starting point we calculate the ensemble average of a product of $l$ SPDOS, namely,
\begin{widetext}
\begin{equation}\label{prelim}
    \begin{split}
    \left\langle\rho\!\left(x_1\right)...\rho\!\left(x_l\right)\right\rangle &\equiv \sum_{n_1,...,n_l} \left\langle \prod_{i=1}^l \delta\!\left(x_i-\varepsilon_{n_i}\right) \right\rangle 
        = \sum_{L_1 \sqcup ... \sqcup L_m = \{1,...,l\}} \sum_{\substack{n_1,...,n_m \\ \forall a \neq b:~n_a \neq n_b}} \left\langle\prod_{i=1}^m \prod_{j \in L_i} \delta\!\left(x_j - \varepsilon_{n_i}\right)\right\rangle \\
        &= \sum_{L_1 \sqcup ... \sqcup L_m = \{1,...,l\}} \sum_{\substack{n_1,...,n_m \\ \forall a \neq b:~n_a \neq n_b}} \lim\limits_{M \to \infty} \int\text{d}e_1 \cdots\!\!\int\text{d}e_M
        \prod_{i=1}^m \prod_{j \in L_i} \delta\!\left(x_j -\rho^{-1}_0\bar{s}_M^{-1}e_{n_i}\right) P_M\!\left(e_1,...,e_M\right).
    \end{split}
\end{equation}
In the first step, we expressed the SPDOS in terms of the set of eigenenergies $\{\varepsilon_{n_i}\}$ of a SP system. 
Next, we reorganize the sums over all eigenvalues $\varepsilon_{n_i}$ in terms of partitions where $|L_i|$ eigenvalues in the product of delta distributions are equal. 
Finally, we use Eq.~\eqref{e_to_eps} to relate the set of eigenenergies of the SP system of interest to the eigenvalues of a random matrix ensemble realization and take the average over the ensemble explicitly.
To relate this expression to Eq. \eqref{convolution} we identify $x_j=E_j/N_j$.

Equation \eqref{prelim} is more conveniently written in terms of the unfolded $m$-point correlation functions $X_m$, namely,
\begin{equation}
\label{def_X_m}
    X_m\!\left(\rho_0\varepsilon_{j_1},...,\rho_0\varepsilon_{j_m}\right) = \lim\limits_{M \to \infty} \bar{s}_M^m \frac{M!}{(M-m)!}  
\int\!\text{d}e_{m+1}\cdots \!\!\int\!\text{d}e_M P_M\!\left(\bar{s}_M\varepsilon_{j_1}\rho_0,...,\bar{s}_M\varepsilon_{j_m}\rho_0,e_{m+1},...,e_M\right),
\end{equation}
which measures the probability density of finding a level around each of $\varepsilon_{j_1},...,\varepsilon_{j_m}$, while the remaining levels are unobserved \cite{mehta2004random,bohigas1991random,Guhr1998}.
Combining Eq.~\eqref{def_X_m} with \eqref{prelim}, we obtain
\begin{equation}
    \label{dos_average_in_terms_of_X}
       \left\langle\rho\!\left(x_1\right)...\rho\!\left(x_l\right)\right\rangle=
        \sum_{L_1 \sqcup ... \sqcup L_m = \{1,...,l\}} \rho_0^m \left(\prod_{i=1}^m \prod_{n=1}^{\left|L_i\right|-1}\delta\!\left(x_{k_n}-x_{k_{n+1}}\right)\right) X_m\!\left(\rho_0x_{j_1},...,\rho_0x_{j_m}\right),
\end{equation}
where we write $L_i = \left\{j_i = k_{i,1},...,k_{i,{\left|L_i\right|}}\right\} = \left\{k_1,...,k_{\left|L_i\right|}\right\}$, omitting the index $i$ unless we wish to emphasize it.
The relation between the correlation functions $X_m$ and the more frequently used cluster functions $Y_n$ reads
\begin{equation}
\label{correlation in terms of clusterfkt's}
    \begin{split}
        X_m\!\left(\rho_0x_{j_1},...,\rho_0x_{j_m}\right) = &\sum_{K_1 \sqcup ... \sqcup K_n = \left\{j_1,...,j_m\right\}} (-1)^{m-n}
        \prod_{i=1}^n Y_{\left|K_i\right|}\!\left(\rho_0 x_j \mid j \in K_i\right).
    \end{split}
\end{equation}
Appendix~\ref{app:cluster_fktS} gives the explicit forms of the 2-point cluster functions $Y^\beta_2$ for the orthogonal $(\beta = 1)$, unitary $(\beta = 2)$ and symplectic $(\beta = 4)$ Gaussian ensembles. 
For general $m$-point cluster functions, we refer the reader to Ref.~\cite{bohigas1991random}.

By combining Eqs.~\eqref{convolution}, \eqref{dos_average_in_terms_of_X}, and \eqref{correlation in terms of clusterfkt's}, we finally arrive at the ensemble-averaged MBDOS of non-interacting indistinguishable particles in terms of the $m$-point cluster functions of the corresponding random matrix ensemble:
\begin{equation}
\label{general_final}
\begin{split}
\left\langle\rho^{(\pm)}(N,E)\right\rangle = &\sum_{l=1}^N (\pm 1)^{N-l} \sum_{\substack{N_1,...,N_l = 1 \\ N_1 \leq ... \leq N_l \\ \sum_{i=1}^l N_i = N}}^N \Tilde{c}\!\left(N_1,...,N_l\right) 
\int_0^\infty \text{d}E_1~... \int_0^\infty \text{d}E_l~ \delta\!\left(E - \sum_{i=1}^l E_i\right)  \\& \times \!\!\!\!\!
\sum_{L_1 \sqcup ... \sqcup L_m = \{1,...,l\}}\!\!\!\!\! \rho_0^m\left(\prod_{i=1}^m \prod_{n=1}^{\left|L_i\right|-1}\delta\!\left(x_{k_n}-x_{k_{n+1}}\right)\right) \sum_{K_1 \sqcup ... \sqcup K_n = \left\{j_1,...,j_m\right\}}
\!\!\!\!\!(-1)^{m-n} \prod_{i=1}^n Y_{\left|K_i\right|}\!\left(\rho_{0}x_j \mid j \in K_i\right) \, ,
\end{split}
\end{equation}
\end{widetext}
recall $x_j = E_j/N_j$. Here, we use the abbreviation
\begin{equation}\label{tilde_c}
\Tilde{c}\!\left(N_1,\dots,N_l\right) = \left[\prod_{N_i \in \left\{N_1,...,N_l\right\}} m\!\left(N_i\right)! 
\prod_{i=1}^l N_i^2\right]^{-1}.
\end{equation} 
This is the main analytical result of this work.

\subsection{Numerical approach}
\label{sec:num_sim}

The evaluation of the ensemble-averaged MBDOS using Eq.~\eqref{general_final}, which involves nested convolutions of high-order cluster functions, presents significant analytical and computational challenges as $N$ increases. 
Analytical solutions are only tractable for specific cases explored in later sections.

To circumvent this limitation, we use a numerical approach that generates the ensemble average
of the MBDOS, defined as
\begin{equation} 
\label{eq:def:numerical-MBDOS}
\rho^{(\pm)}(N, E) = \sum_{\nu} \delta\left(E - E_\nu\right) \, .
\end{equation}
Here $\nu$ labels the MB state configuration defined by the tuple $(n_{i,\nu}^{(\pm)})$ that describes the SP state occupation numbers $n_{i,\nu}^{(\pm)}$, namely, $n_{i,\nu}^{(-)} = 0,1$ for fermions and $n_{i,\nu}^{(+)}= 0, 1,..., N$ for bosons, with
\begin{equation}
    N = \sum_i n_{i,\nu}^{(\pm)} \quad \mbox{and} \quad 
    E_\nu = \sum_i n_{i,\nu}^{(\pm)} \varepsilon_i .
 \end{equation}

The numerical simulation method consists of two steps. 
First, we generate $N_{\rm R} \gg 1$ SP spectra.
For chaotic systems, they are obtained by diagonalizing different realizations of random matrices with dimensions $M\times M$ of a given symmetry class $\beta$. 
The unfolded eigenvalues are obtained by evaluating the integrated semi-circle level density of the random matrices \cite{mehta2004random}. 
Next, we introduce lower and upper energy cutoffs, $\varepsilon_0$ and $\varepsilon_{\rm max}$, to discard the extreme eigenvalues whose fluctuations are strongly affected by the eigenvalue distribution, the so-called confining potential \cite{mehta2004random}. 
Here, for every ensemble realization, we choose $\varepsilon_0$ and $\varepsilon_{\rm max}$ to discard at least the $50$ lowest and highest eigenvalues
and use the shifted SP energy spectrum $[\varepsilon_1,\varepsilon_2,\dots]\rightarrow [\varepsilon_1-\varepsilon_0,\varepsilon_2-\varepsilon_0,\dots]$ for the numerical simulations.
In contrast, for integrable systems, the spectra are constructed by a cumulative summation of random variables drawn from a Poisson distribution.

We proceed by systematically enumerating all possible SP occupation configurations to obtain the set of tuples, denoted by $\{(n_i^{(\pm)})\}$, over which the sum in Eq.~\eqref{eq:def:numerical-MBDOS} is carried out.
We then populate the SP levels according to these configurations.
In practical terms, for each realization $r$ of the SP spectrum, we obtain the corresponding MB spectrum by considering all $\nu$ tuples, $(n_{i,\nu}^{(\pm)})$, from the set $\{(n_i^{(\pm)})\}$.
As a result, the MB energies are given by $E_{\nu,r} = \sum_i n_{i,\nu}^{(\pm)} \varepsilon_{i,r}^{}$.
Finally, the ensemble average is obtained by averaging the calculated MB density of states, $\rho_r^{(\pm)}(N, E)$, over all SP spectrum realizations.

While calculating the ensemble average $\langle\rho^{(\pm)}(N,E)\rangle$  through generation and population of SP spectra is possible, we find it more convenient to use the energy integrated MBDOS, also known as counting function, $\mathcal{N}^{(\pm)}(N, E)$ instead.  
For a given SP spectrum realization $r$, 
$\mathcal{N}^{(\pm)}_r(N, E)$ counts the number of MB states $t$ for which $\sum_i n^{(\pm)}_{i,\nu} \varepsilon_{i,r}\leq E$ with the constraint $\sum_i n^{(\pm)}_{i,\nu} = N$.

Calculations using $\mathcal{N}^{(\pm)}(N, E)$ avoid the numerical difficulties associated with delta functions and are computationally amenable for analyzing data obtained from numerical simulations. 
Hence, we employ $\mathcal{N}^{(\pm)}(N, E)$ for the numerical analysis presented in this work.

Before presenting our results, a technical comment is in order. 
Based on the Weyl expansion \cite{Berry1994}, one expects that both $\langle\rho^{(\pm)}(N,E)\rangle$ and $\langle\mathcal{N}^{(\pm)}(N,E)\rangle$ will be well approximated by a polynomial expansion with descending powers of $E$. 
Therefore, the choice of the energy of the lowest SP level $\varepsilon_1$ affects all subleading orders of the MBDOS expansion. 
Hence, to compare the numerical results with the analytical ones of Eq.~\eqref{general_final}, 
the cutoff $\varepsilon_0$ 
has to be uncorrelated with all subsequent energy levels $[\varepsilon_1,\varepsilon_2,\dots]$ in order to be compatible with the RMT assumptions; see the discussion in App.~\ref{app:first_level}.
There, we show that the distribution of $\varepsilon_1$ is given by
\begin{equation}\label{eq:first_level_stat}
    p_1^{(\beta)}(\varepsilon_1) = \rho_0 \int_0^\infty \text{d}a\  p^{(\beta)}(a+\varepsilon_1),
\end{equation}
where $p^{(\beta)}$ is the RMT level-spacing distribution \cite{mehta2004random}. 
Note that the probability of placing $\varepsilon_0$ between two SP energy levels grows with their spacing. 
It follows that the expectation value of the $\varepsilon_1$ is larger than $\rho_0^{-1}/2$. 
For SP spectra with Poisson statistics, the distribution of the first SP level again follows a Poisson distribution.

Therefore, the mean MB ground state energy $\langle E_{\rm GS} \rangle$
for the SP spectra generated this way differs for each class of ensemble. 
Thus, whenever we compare $\langle\mathcal{N}^{(\pm)}(N,E)\rangle$ for different ensembles, we shift the curves such that all ensembles are displayed at the same {\it average} excitation energy $\langle Q \rangle=E-\langle E_{\rm GS} \rangle$.

In Sec.~\ref{sec:signatures}, we build and analyze the ensemble average of an apparently similar object, $\mathcal{N}^{(\pm)}(N,Q)$, the cumulative MBDOS as a function of the excitation energy, $Q=E-E_{\rm GS}$. 
Since $E_{\rm GS}$ shows a large dispersion over the ensemble, we show that $\langle \mathcal{N}^{(\pm)}(N,Q)\rangle$ and $\langle \mathcal{N}^{(\pm)}(N,E)\rangle$
differ, in general, significantly.

\section{Mean MBDOS of non-interacting systems with integrable SP dynamics}
\label{sec:results_integrable}

The SP spectra of systems with integrable dynamics are known to exhibit Poisson statistics \cite{Stöckmann_1999,haake2001}, which implies uncorrelated SP energy level spacings.
However, even in this limit, spectral correlations emerge at the MB level due to the fact that all $N$ particles of the gas share the same SP spectrum. 
In what follows, we will single out this cumulative effect arising from the combinatorial contribution of (uncorrelated) SP level fluctuations to the ensemble-averaged MBDOS.
We refer to this feature as (induced) MB spectral correlations to be distinguished from SP level correlations of quantum chaotic SP systems.

For Poisson statistics, the joint eigenvalue probability density $P_M$ factorizes. 
As a consequence, the unfolded $m$-point correlation functions become an $m$-fold product of the constant probability density 
of finding a level at some position in the unfolded spectrum (choosing $\lim_{M\to\infty}\bar{s}_M=\bar{s}=1$):
\begin{equation}\label{X_for_Poisson}
    X_m\!\left(\rho_{0}x_1,...,\rho_{0}x_m\right) = \prod_{i=1}^m \bar{s}^{-1} = 1.
\end{equation}
Thus, by combining Eqs.~\eqref{convolution}, \eqref{dos_average_in_terms_of_X}, and \eqref{X_for_Poisson}, the ensemble-averaged MBDOS is reduced to the simple form
\begin{widetext}
\begin{equation}\label{poisson_prelimiary}
    \begin{split}
        \left\langle\rho^{(\pm)}(N,E)\right\rangle = \sum_{l=1}^N (\pm 1)^{N-l} \sum_{\substack{N_1,...,N_l = 1 \\ N_1 \leq ... \leq N_l \\ \sum_{i=1}^l N_i = N}}^N \Tilde{c}\!\left(N_1,...,N_l\right) 
        \int_0^\infty &\text{d}E_1~... \int_0^\infty \text{d}E_l~\delta\!\left(E - \sum_{i=1}^l E_i\right)\\
        \times &\!\!\!
        \sum_{L_1 \sqcup ... \sqcup L_m = \{1,...,l\}} \left(\prod_{i=1}^m \prod_{n=1}^{\left|L_i\right|-1}\delta\!\left(x_{k_n}-x_{k_{n+1}}\right)\right)\rho_0^m,
    \end{split}
\end{equation}
where all integrals can now be evaluated explicitly to yield the following polynomial expression in energy (see App.~\ref{Poisson_integration} for more details)
\begin{equation}\label{final_poisson}
    \begin{split}
        \left\langle\rho^{(\pm)}(N,E)\right\rangle = \frac{1}{N!}\sum_{l=1}^N (\pm 1)^{N-l} \!\!\!\!\!\!
    \sum_{\substack{N_1,...,N_l = 1 \\ N_1 \leq ... \leq N_l \\ \sum_{i=1}^l N_i = N}}^N  \!\!\!\!\!\! c \left(N_1,...,N_l\right) \!\!\!\!\!\!\!\!
        \sum_{L_1 \sqcup ... \sqcup L_m = \{1,...,l\}} \!\!\!\!\!\!\!\!
        c_{L_1,...,L_m}\!\left(N_1,...,N_l\right) \frac{E^{m-1}}{(m-1)!}\rho_0^m,
    \end{split}
\end{equation}
\end{widetext}
where
\begin{equation}\label{poisson_coefs}
    c_{L_1,...,L_m}\!\left(N_1,...,N_l\right) = \prod_{i=1}^m \left[\sum_{k \in L_i} N_k\right]^{-1}.
\end{equation} 

Let us begin the analysis of Eq.~\eqref{final_poisson}, focusing on the disjoint partitions of the set $\{1,..,l\}$ fulfilling $|L_i|=1$.

Since this is in exact correspondence with
neglecting MB spectral correlations arising from SP level fluctuations in Eq.~\eqref{convolution},
we infer that the resulting contribution to the MBDOS, which we call $\bar{\rho}^{(\pm)}$, accounts for indistinguishability, but not for induced MB spectral correlations.
Accordingly, $\bar{\mathcal{N}}^{(\pm)}$ is expected to grow approximately at each mean MB energy level $\langle E_\nu\rangle$.

A similar result was obtained in Ref. \cite{Quirin} using a semiclassical analysis.
The latter analyzed the Laplace transform of the traced (anti-) symmetrized MB propagator in the limit of short-time paths. 
For non-interacting particles, the MB propagator factorizes and SP periodic orbits corrections are suppressed in the limit of short-time paths. 
As a consequence, this approximation only captures the smooth part of the SPDOS, and the agreement with $\bar{\rho}^{(\pm)}$ is not surprising. 
Reference~\cite{Quirin} demonstrated that  $\bar{\rho}^{(\pm)}$ corresponds to the smooth part of the MBDOS for an effective two-dimensional finite system without boundaries, where the SP mean level density is constant. When comparing $\bar{\rho}^{(\pm)}$ with $\langle\rho^{(\pm)}\rangle$ or their associated counting functions, we need to consider the distinct (mean) SP ground state energies, as explained in Sec.~\ref{sec:num_sim}. This adjustment is important
to account for zero-point contributions to the energy as, for example, in the harmonic oscillator. 

The remaining disjoint partitions ($|L_i|\neq1)$ in Eq.~\eqref{final_poisson} comprise the so-called contact terms. 
They affect all orders of the energy power expansion, except the leading one. 
One can understand their contributions to originate from both indistinguishability and induced MB spectral correlations.

For boson gases, the evaluation of $\langle\rho^{(+)}(N,E)\rangle$ using Eq.~\eqref{final_poisson} is straightforward, although somewhat messy. 
In Appendix \ref{app:poisson_bososns} we present the explicit results for a few values of $N$, namely $N=2,3,5,\text{ and }8$.

In the fermionic case, we find, through a rather convoluted algebra, that all subleading contributions cancel.
Consequently, the ensemble-averaged MBDOS is reduced to the Weyl volume term for a symmetrized phase space volume, namely,   
\begin{equation}\label{WV_poisson}
    \left\langle\rho^{(-)}(N,E)\right\rangle=\frac{E^{N-1}}{N!(N-1)!}\rho_0^N.
\end{equation}
A semi-analytical derivation of this result is presented in Appendix~\ref{app:poissom_fermions}, where we introduce a set of modified Stirling numbers and numerically verify Eq.~\eqref{WV_poisson} by analyzing Eq.~\eqref{final_poisson}  up 
to $N=10$.

In contrast, Ref.~\cite{Quirin} finds an oscillatory behavior for $\bar{\rho}^{(-)}(N,E)$ at low energies ($E \alt \langle E_\text{GS}\rangle$), due to the alternating signs of subleading terms.
The slow growth at low energies is attributed to a manifestation of the Fermi energy. 
Equation \eqref{WV_poisson} does not display such oscillations, indicating that the ensemble average over SP spectra with Poisson statistics suppresses the signatures of a Fermi surface. 

Notably, the first sub-leading contribution to $\bar{\rho}^{(\pm)}$ in the $E$-expansion has the same magnitude as the first correction term arising from both indistinguishability and induced MB spectral correlations. 
Unlike the fermionic case, both contributions have the same sign for bosons. 
Consequently, for both particle statistics, the subleading contribution to  
$\langle\rho^{(\pm)}\rangle$ is amplified compared to $\bar{\rho}^{(\pm)}$, but it is twice as large in the bosonic case.
 
\begin{figure}
     \centering
     \includegraphics[width=0.9\linewidth]{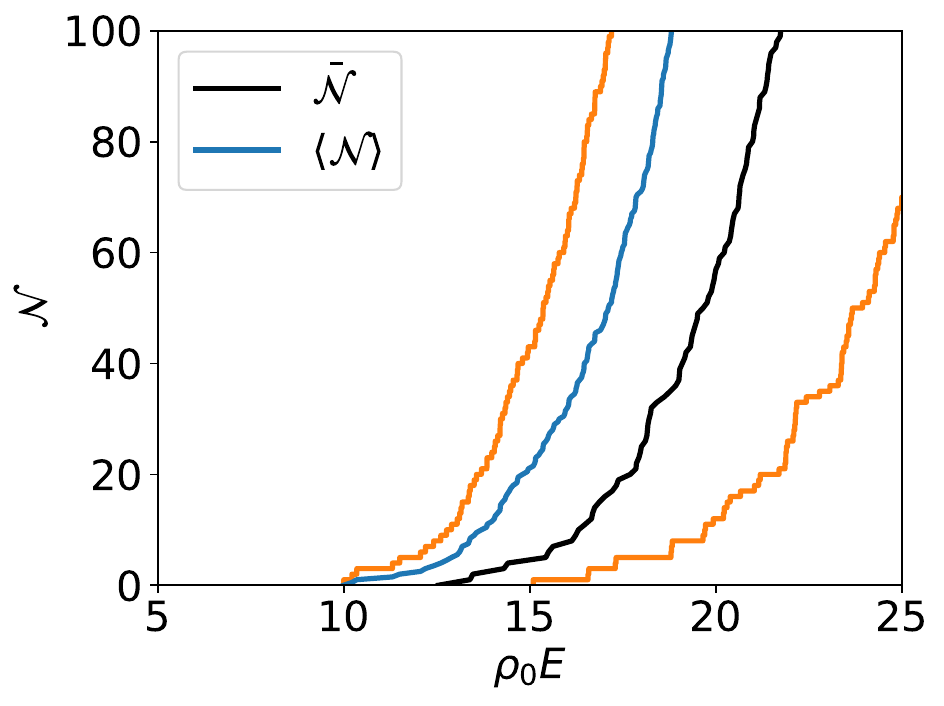}
     \caption{
     Cumulative fermionic MBDOS $\mathcal{N}^{(-)}(E, N)$ of two ensemble realizations (orange) with different ground state energies $E_{\rm GS}$ and their mean $\langle\mathcal{N}^{(-)}\rangle$ (blue) for $N=5$.
     The counting function following the average level position (black) of the two realizations, $\bar{\cal N}^{(-)}$ (introduced below Eq.~(\ref{poisson_coefs})) is also depicted.
     }
     \label{fig:example_average}
\end{figure}

Let us now switch to the analysis of the numerical simulations. 
Figure~\ref{fig:example_average} serves as a guide to interpreting the results.
It displays the cumulative MBDOS, $\mathcal{N}^{(-)}(E, N=5)$, of two representative ensemble
realizations (orange) together with their mean $\langle\mathcal{N}^{(-)}\rangle$ (blue). 
The black line shows an alternative way to take the average, namely, the counting function following the mean position of each MB level within the ensemble.
Figure~\ref{fig:example_average} clearly shows that the growth of $\langle\mathcal{N}^{(-)}\rangle$ at low energies, here $\rho_0 E \alt 15$, is dominated by the realization with the lowest $E_\text{GS}$.
Figure \ref{fig:example_average} also suggests that the growth onset of $\langle\mathcal{N}^{(-)}\rangle$ is dominated by the excited states of the ensemble realizations with $E_{\rm GS} < \langle E_\text{GS}\rangle$.
Unlike $\langle\mathcal{N}^{(-)}\rangle$, the growth onset of $\overline{}{\mathcal{N}}^{(-)}$ 
coincides with the mean fermionic ground state energy $\langle E_\text{GS}\rangle$. 
As a consequence, $\langle\mathcal{N}^{(-)}\rangle$ is expected to contain little information about the Fermi surface.

\begin{figure}
    \centering
    \includegraphics[width=0.9\linewidth]{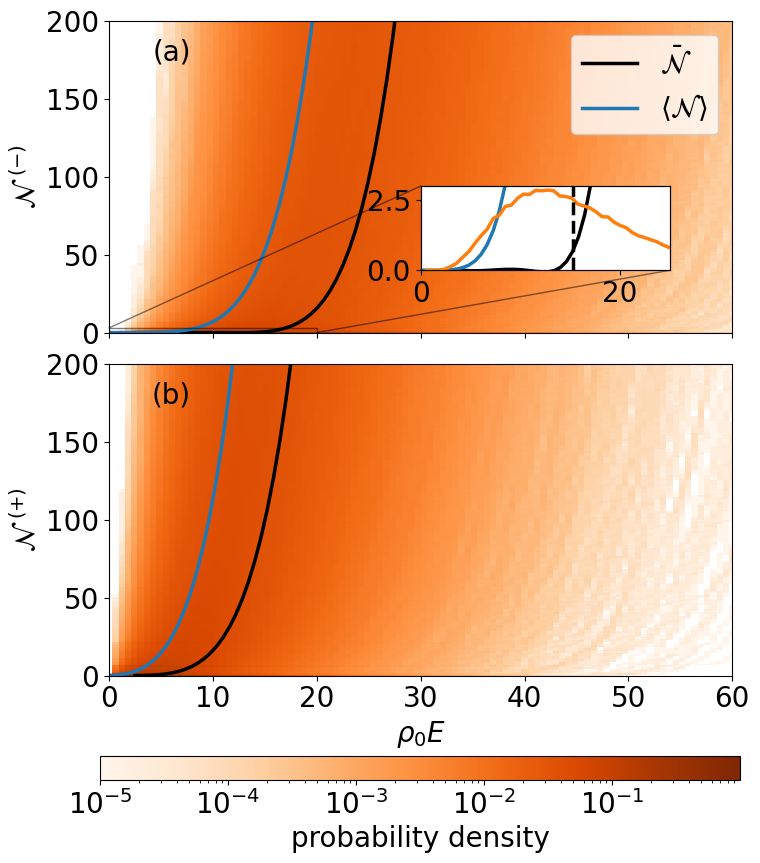}
    \caption{
        Cumulative MBDOS of (a) fermionic ${\cal N}^{(-)}(E, N)$ and (b) bosonic  ${\cal N}^{(+)}(E, N)$ systems with $N=5$ non-interacting particles with integrable SP dynamics.
        The color intensity indicates the probability density $P({\cal N}^{(\pm)})$ computed for an ensemble of $N_{\rm R} = 10^5$ realizations. 
        The blue line stands for the ensemble averaged $\langle\mathcal{N}^{(\pm)}(E, N)\rangle$ and the black one for $\bar{\mathcal{N}}^{(\pm)}(E, N)$. 
        Inset: Magnification at small values of the average cumulative MBDOS $\langle\mathcal{N}\rangle$ and $\bar{\mathcal{N}}$, together with the probability density $P(E_{\rm GS})$ of finding a realization with a ground state energy $E_{\rm GS}$ 
        (in orange, arbitrary units). The dashed black line indicates $\langle E_{\rm GS}\rangle$.
        }
        \label{GS_study}
\end{figure}

Figure~\ref{GS_study} illustrates the large fluctuations of ${\cal N}^{(\pm)}(E, N)$
for both (a) fermionic and (b) bosonic systems with integrable SP dynamics. The results correspond to an ensemble of systems with $N=5$ particles with $N_{\rm R} = 10^4$ realizations. 
In the fermionic case, these fluctuations can be partially attributed to large variations of the ground state energy $\langle E_{\rm GS}\rangle$, shown in the inset of Figure \ref{GS_study}(a).

Figure~\ref{GS_study}(b) shows that the variance of the cumulative MBDOS is also large for bosons. 
Similar to the fermionic case, $\langle\mathcal{N}^{(+)}\rangle$ exceeds $\bar{\mathcal{N}}^{(+)}$ as a consequence of a fraction of realizations growing much faster with $E$ than  $\bar{\mathcal{N}}^{(+)}$. 

\section{Mean MBDOS of non-interacting systems with chaotic SP dynamics} 
\label{sec:Chaos}

Let us now study the specific impact 
of spectral correlations between SP levels, obeying Gaussian RMT statistics, on $\langle\rho^{(\pm)}(N,E)\rangle_\beta$. 
In contrast to the Poisson statistics,
SP level correlations are the hallmark of quantum chaos. 
As a consequence, the correlation functions in Eq.~\eqref{correlation in terms of clusterfkt's} exhibit a complex structure and depend on the system's universality class. 
The latter can be expressed in terms of cluster functions $Y_m(x_1,...,x_m)$, as in Eq.~\eqref{general_final}.
For $N=2$ we can analytically calculate the convolution integrals and conveniently decompose the MBDOS as
\begin{equation}\label{DOS_RMT_N2}
    \begin{split}
        \langle\rho^{(\pm)}(2,E)\rangle_\beta=\langle\rho^{(\pm)}(2,E)\rangle_P\ +\phantom{JD loves Rick Astley}&\\
        \begin{cases}
            \frac{\rho_0}{4}+\frac{1+\text{Si}(\pi E\rho_0)\sin(\pi E\rho_0)-\cos(2\pi E\rho_0)}{2\pi^2 E}\\
            \;\;\;\; -\frac{\sin(\pi E\rho_0)}{4\pi E}-\frac{\text{Si}(2\pi E\rho_0)}{\pi}\rho_0, &\beta=1\\ 
            \frac{1-\cos(2\pi E\rho_0)}{4\pi^2E}-\frac{\text{Si}(2\pi E\rho_0)}{2\pi}\rho_0, &\beta=2\\ 
            \frac{1+\text{Si}(2\pi E\rho_0)\sin(2\pi E\rho_0)-\cos(4\pi E\rho_0)}{8\pi^2 E}\\
            \;\;\;\; -\frac{\text{Si}(4\pi E\rho_0)}{2\pi}\rho_0, &\beta=4
        \end{cases}
    \end{split}
\end{equation}
where the first term,
\begin{equation}
    \langle\rho^{(\pm)}(2,E)\rangle_P=E/2\pm1/4+1/4 \, , 
\end{equation}
involves an ensemble average $\langle \cdots \rangle_P$ over SP spectra with Poisson statistics. It includes the leading Weyl volume term, corrections due to indistinguishability, and contributions due to MB spectral correlations from Poisson-type SP energy level fluctuations, as discussed in the previous section. 
Notably, for fermions the latter two terms precisely cancel and only the leading Weyl term remains. 

\begin{figure}
    \centering
    \includegraphics[width=\linewidth]{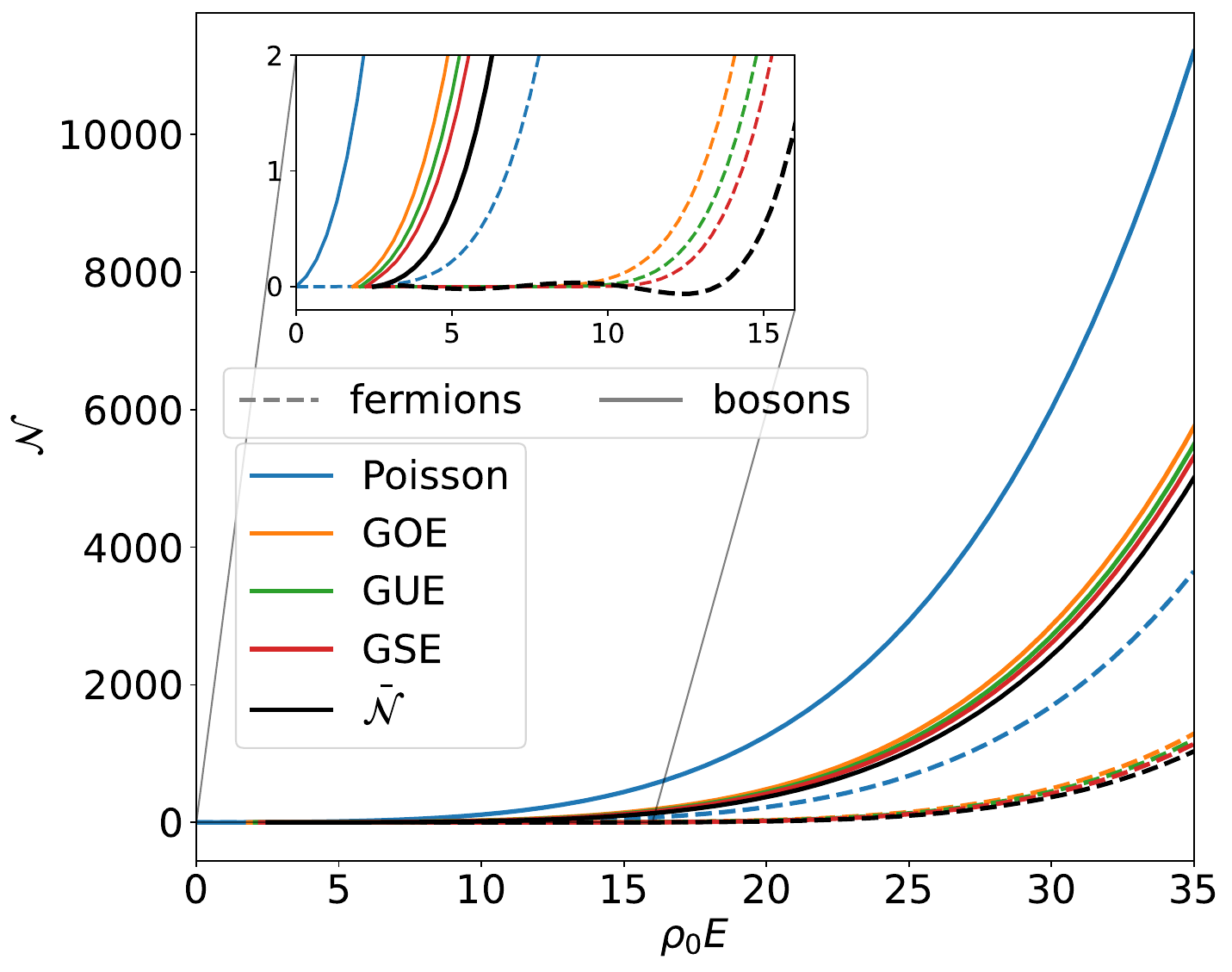}
    \caption{
    Ensemble-averaged cumulative MBDOS $\langle\mathcal{N}\rangle$ as a function of the energy for $N=5$ fermions (dashed line) and bosons (solid line) for the following cases: Poisson: blue, GOE ($\beta=1$): orange, GUE ($\beta=2$): green, GSE ($\beta=4$): red. 
    The inset shows a magnification of the low energy regime, indicating that for increasing level repulsion, the integrated MBDOS approaches the limit of a MBDOS arising from a SP spectrum with a constant level spacing, $\bar{\mathcal{N}}$ (black line). 
    In the case of chaotic SP systems the ensemble consisted of $N_{\rm R} = 10^5$ realizations while for the integrable case we used Eqs.~\eqref{poissonN2358} and \eqref{WV_poisson}.}
    \label{Comparison_N2_counting}
\end{figure}

The second term in Eq. \eqref{DOS_RMT_N2} directly reflects the effect of random-matrix correlations between SP levels on the ensemble-averaged MBDOS (depending on the respective universality class).

In the limit $\rho_0 E \gg 1$,  Eq.~\eqref{DOS_RMT_N2} yields
\begin{gather}
    \langle\rho^{(\pm)}(2,E)\rangle_\beta\stackrel{\rho_0 E \gg 1}{=}\langle\rho^{(\pm)}(2,E)\rangle_P\ -\frac{\rho_0}{4}
\end{gather}
for all three symmetry classes,  $\beta = 1, 2$, and 4. We conclude that RMT-type SP correlations reduce the average MB density of states. 

For $N>2$, the integrals in Eq.~\eqref{correlation in terms of clusterfkt's} become so involved that we find it more convenient to solve them using numerical methods or to rely on numerical simulations, as introduced in Sec.~\ref{sec:num_sim}. 
In Appendix~\ref{app:error-analysis} we discuss the precision of these methods.

Figure~\ref{Comparison_N2_counting} shows the dependence of the cumulative MBDOS on the symmetry class, and hence spectral rigidity, of the underlying SP spectrum.
With increasing SP level repulsion, i.e.,  increasing $\beta$, the ensemble-averaged MB staircase function approaches the limit of a cumulative MBDOS $\bar{\mathcal{N}}^{(\pm)}$, based on a SP spectrum without fluctuations. This trend, illustrated in Figure \ref{Comparison_N2_counting} for $N=5$ particles, is already found for $N=2$ particles, and also holds as $N$ increases.

In contrast, for systems with underlying integrable SP motion, we have demonstrated in Sec.~\ref{sec:results_integrable} that the ensemble-averaged MBDOS differs from $\bar{\mathcal{N}}^{(\pm)}$ in all, but the leading order in $E$. These findings indicate that the signatures of the SP dynamics, whether integrable or chaotic, are encoded in the ensemble-averaged MBDOS at each sub-leading order in an energy expansion.
We note that the discrepancy between the ensemble-averaged MBDOS for systems with integrable and chaotic SP dynamics grows not only with $E$ at a fixed $N$, as shown in Fig. \ref{Comparison_N2_counting}, but also with $N$ at a fixed $E$ (not shown here).

\begin{figure}
    \centering
    \includegraphics[width=0.9\linewidth]{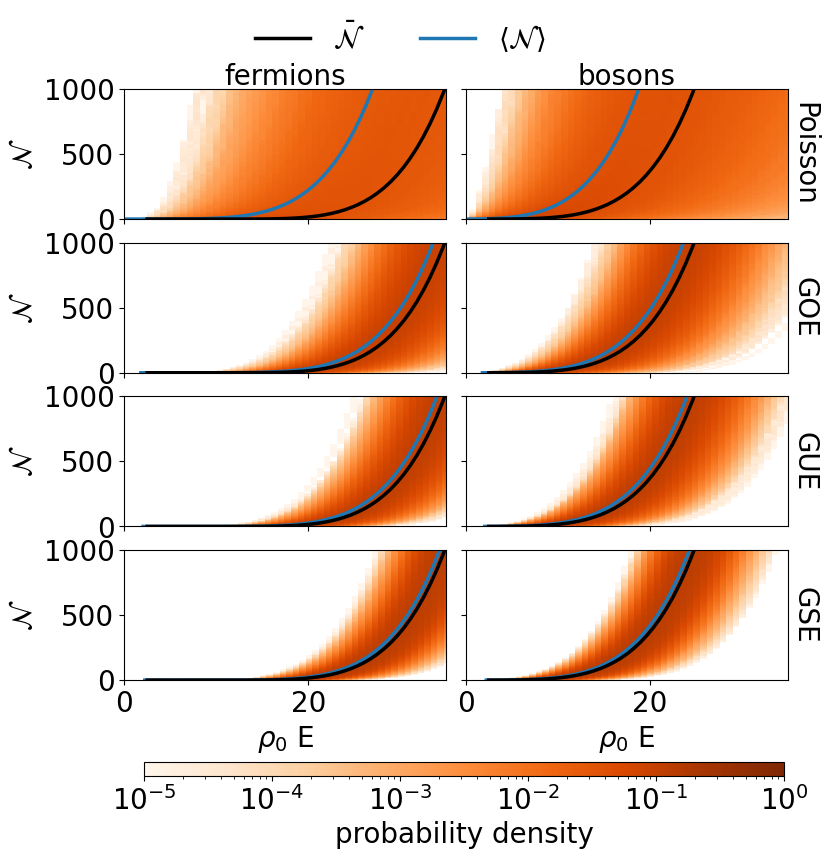}
    \caption{
    Probability density (see color bar) for the occurrence of a MB staircase function obtained from $10^5$ realizations for $N=5$ non-interacting fermions (left column) and bosons (right column) with underlying integrable (Poisson) and chaotic (GOE, GUE, GSE) SP dynamics (from top to bottom). For reference, the ensemble averaged MB counting function $\langle\mathcal{N}\rangle$ (blue) and $\bar{\mathcal{N}}$ (black) are also depicted.
    }
    \label{Variance_of_all_ensembles}
\end{figure}

As discussed in Sec.~\ref{sec:results_integrable}, the rapid growth of the MB counting function and MBDOS at energies below the mean ground state energy for systems with integrable SP dynamics, is due to the large variances characteristic to the Poisson ensemble and their effect on the ensemble average. 
For the MB ensembles with chaotic SP dynamics, the variance decreases with increasing SP level repulsion, as shown in Figure \ref{Variance_of_all_ensembles}.

Usually, the ergodic property asserts the equivalence of ensemble averages and running (energy) averages, provided the Hamiltonian is drawn from a fully random matrix ensemble, for both the fermionic and bosonic cases.
While ergodicity has been proven to hold, with some notable exceptions \cite{Asaga2001}, even for embedded ensembles representing systems with $k$-body interactions (see, for instance, Ref.~\cite{Kota2018} for a review) the results from this and the previous section support our claim of a strong breaking of ergodicity for ensembles of non-interacting MB systems. MB systems of independent particles and mean-field descriptions of interacting MB systems exhibit strong deviations from full RMT. 

We observe that this lack of ergodicity manifests itself at two levels. 
First, when considered as a function of the bare total energy $E$, we observe how the realization-to-realization fluctuations of the MB ground state energy effectively scramble the sub-leading contributions to the MB level density due to indistinguishability.
This effect is further reinforced in the integrable (Poissonian) case, to the extent that only the leading large-$E$ behavior of the fermionic density of states is visible after ensemble averaging, in accordance with Eq.~\eqref{WV_poisson}.

\begin{figure}
    \centering
    \includegraphics[width=0.85\linewidth]{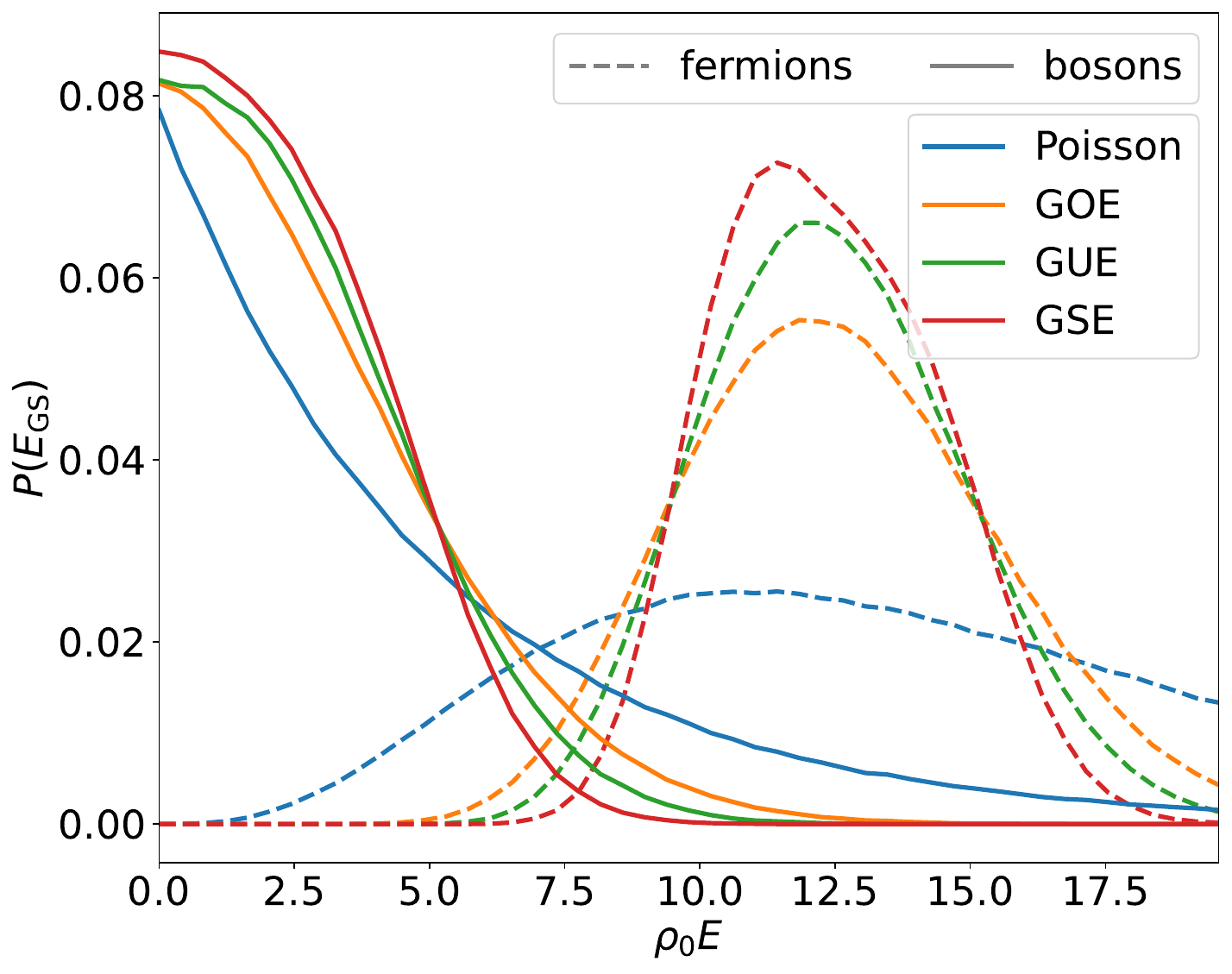}
    \caption{Probability distribution $P(E_{\rm GS})$ of the MB ground state energy $E_{\rm GS}$ for $N=5$ non-interacting bosons (solid line) and fermions (dashed line), for the following cases: Poisson: blue, GOE: orange, GUE: green, GSE: red. The data is obtained from an ensemble comprising $N_R=10^6$ realizations.}
    \label{fig:P(E_GS)}
\end{figure}

Second, the large fluctuations observed in Fig.~\ref{Variance_of_all_ensembles} are partially due to the dispersion of the ground state energies $E_{\rm GS}$ across different realizations.
Figure \ref{fig:P(E_GS)} shows the probability distribution of the MB ground state energy for $N=5$ fermions and bosons. 
For bosons, the MB ground state energy is, as to be expected, related to the SP first-level statistics, detailed in App.~\ref{app:first_level}. 
In contrast, the fermionic case shows a pronounced spread in the MB ground state energy for integrable SP systems, while a much narrower distribution is observed in the RMT case.

We note that even when these MB ground state fluctuations are removed by measuring the energies for each realization relative to the MB ground state, i.e., in terms of the MB excitation energy $Q$, the fluctuations are reduced only in the chaotic case. 
 This will be discussed in the next section.

\section{Signatures of chaotic and integrable SP dynamics in the excitation energy MBDOS}
\label{sec:signatures}

\subsection{{Mean integrated MBDOS versus ensemble fluctuations}}

A recent work \cite{Echter2024} has shown that the MBDOS as a function of excitation energy, $Q = E - E_{\rm GS}$, for systems of indistinguishable non-interacting particles with {\em equally spaced} SP spectra does not distinguish between fermions and bosons. 
Furthermore, notably, there is a regime where the MBDOS for a given $Q$ is independent of the system's number of particles. 
Given the contrast to the findings presented in the previous sections, a natural question arises: How general are these results?
In other words, what happens for SP spectra with {\em fluctuating} SP energy level spacings? 

Importantly, experimental setups  typically involve probing a single or a few systems at a fixed excitation energy $Q$ or varying a control parameter across a single or a few realizations.
Hence, since the mean $\langle\mathcal{N}^{(\pm)}(N,Q)\rangle$ is only a meaningful quantity when $\langle\mathcal{N}^{(\pm)}\rangle\gg\delta \mathcal{N}^{(\pm)}$ it is important to investigate the standard deviation $\delta\mathcal{N}(N,Q)$ within a given ensemble of realizations. 

\subsection{{Dependence on excitation energy $Q$}}

In this subsection, we hence analyze $\langle\mathcal{N}^{(\pm)}(N,Q)\rangle$ and $\delta\mathcal{N}^{(\pm)}(N,Q)$ as a function of excitation energy $Q$ and demonstrate its surprisingly large sensitivity on the underlying SP dynamics.
Here, our conclusions are drawn from extensive numerical simulations, since neither $\langle{\cal N}^{(\pm)}(N,Q)\rangle$ nor $\delta\mathcal{N}^{(\pm)}(N,Q)$ are  amenable quantities to calculate analytically.

The $\langle\rho^{(-)}(N,Q)\rangle$ (or $\langle{\cal N}^{(-)}(N,Q)\rangle$) can be significantly different from $\langle\rho^{(+)}(N,E)\rangle$ (or $\langle{\cal N}^{(+)}(N,E)\rangle$), since the fermionic $E_{\rm GS}$ can exhibit large fluctuations, as illustrated by Figure~\ref{fig:P(E_GS)}.
For instance, as discussed in Sec.~\ref{sec:results_integrable}, the absence of subleading terms in the ensemble-averaged MBDOS for fermionic systems $\langle\rho^{(-)}(N,E)\rangle$ with underlying integrable SP dynamics can be attributed to the dominance of realizations with low MB fermionic $E_{\rm GS}$ (compared to the mean) on the ensemble average. 

\begin{figure}
    \centering
    \includegraphics[width=\linewidth]{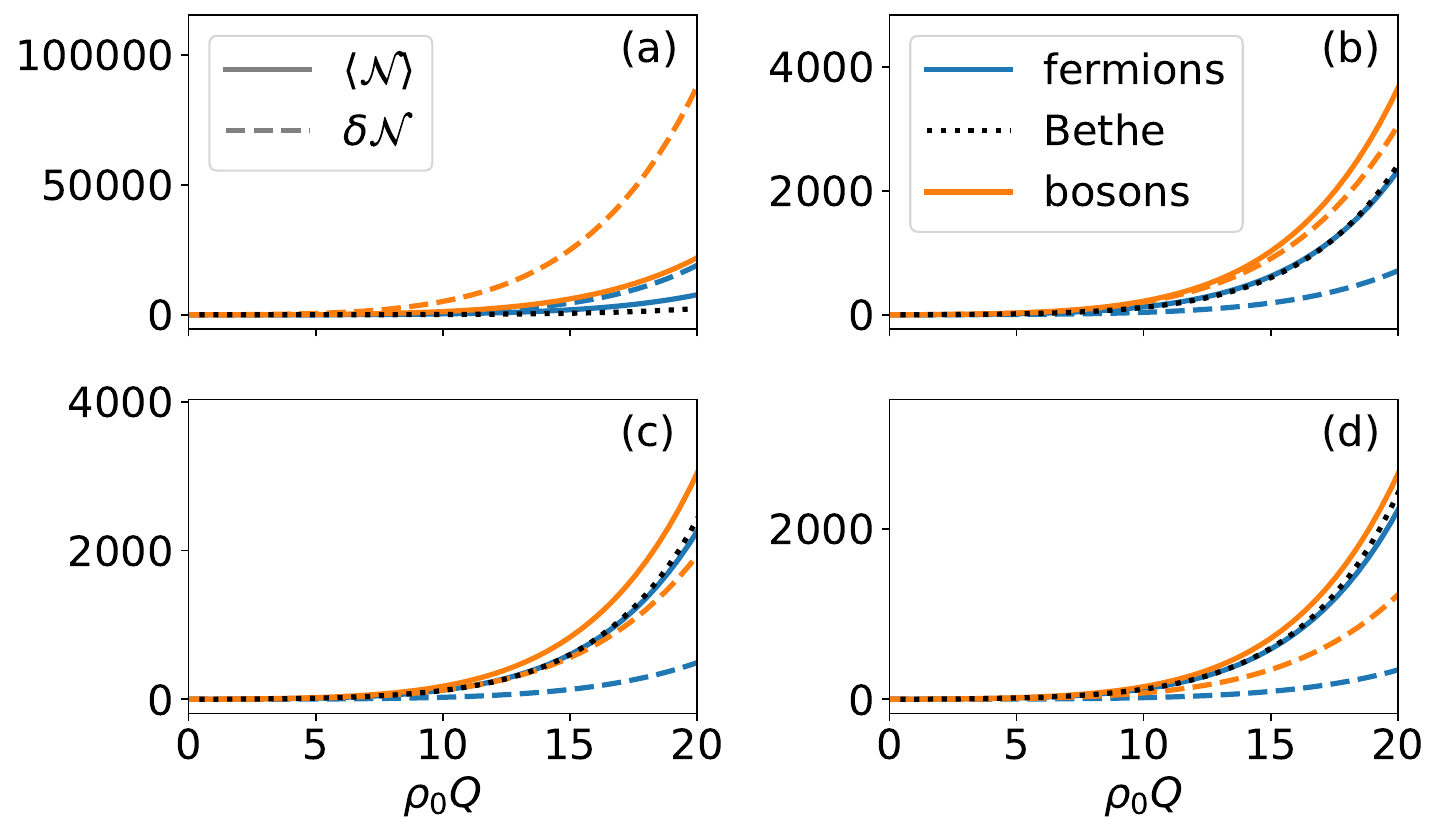}
    \caption{The ensemble-averaged MB counting function (solid) for $N=10$ fermions (blue) and bosons (orange) as well as the standard deviation (dashed) within the ensembles are shown. 
    Each realization within the ensemble is measured at its excitation energy $Q$ (difference between total energy E and its MB ground state energy). 
    The shown ensembles comprise SP spectra of Poisson (a), GOE (b), GUE (c) and GSE (d) statistics. Each ensemble contains $N_{\rm R} = 10^5$ realizations.}
    \label{std_for_each_ensemble}
\end{figure}

Figure \ref{std_for_each_ensemble} shows both $\langle\mathcal{N}(N,Q)\rangle$ (solid line) and $\delta\mathcal{N}(N,Q)$ (dashed line) for systems with Poisson, GOE, GUE, and GSE single-particle energy level statistics. 
The ensemble averages comprise $N_{\rm R} = 10^5$ realizations.
For bosonic and fermionic MB systems with underlying integrable SP dynamics, the mean $\langle\mathcal{N}(N,Q)\rangle$ is clearly not a representative statistical measure, since $\langle\mathcal{N}(N,Q)\rangle\ll\delta\mathcal{N}(N,Q)$.
Moreover, fluctuations in bosonic systems consistently exceed those for fermions. 
Interestingly, for the latter, even the linear level repulsion characteristic of 
GOE SP spectra is sufficient to significantly reduce the fluctuations, that is, $\langle\mathcal{N}(N,Q)\rangle\gg\delta\mathcal{N}(N,Q)$. 
This explains the success of the Bethe formula  \cite{Bethe1936}, an asymptotic expression for the fermionic MBDOS for large $N$ and $Q$, which, by construction, captures only  the smooth part of the SPDOS. 

\subsection{{Dependence on particle number $N$}}

\begin{figure}
    \centering
    \includegraphics[width=\linewidth]{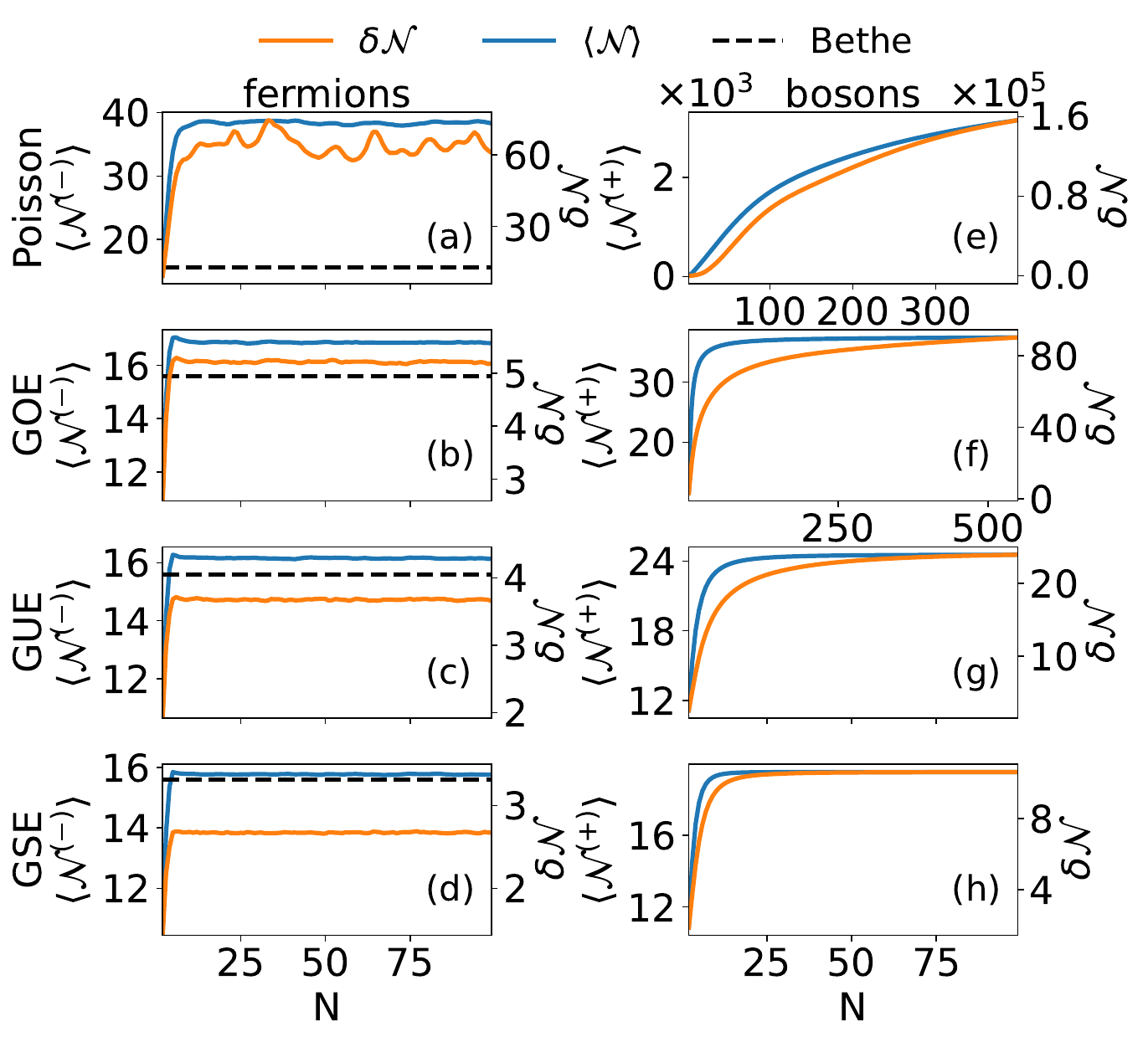}
    \caption{Ensemble averaged MB counting function $\langle\mathcal{N}(N,Q)\rangle$ (blue, left $y$-axis) and its variance $\delta\mathcal{N}$ (orange, right $y$-axis) as a function of the particle number $N$ for an excitation energy of $\rho_0 Q=5$ and $N_{\rm R} = 10^5$ realizations. 
    The results for fermion/boson systems are displayed in the left/rights column ((a)-(d))/((e)-(h)), while the rows represent ensembles of Poisson, GOE, GUE, and GSE single-particle statistics.
    Within the Poisson ensemble, realizations with degenerate SP GS where excluded.
    The Bethe formula for $\langle\mathcal{N}(N,Q)\rangle$ is depicted by a black dashed line for the fermionic cases ((a)-(d)). 
    For bosons with underlying Poisson (e) and GOE (f) SP energy level statistics, we show results for $N\le 400$ and $N\le 550$ particles, while for all other cases $N\le 100$.
    }
    \label{fig:eff_D}
\end{figure}

Having analyzed $\mathcal{N}(N,Q)$ for fixed values of $N$, we now turn our attention to the case of fixed $Q$ and varying $N$. 
We note that a similar study has been conducted in Ref.~\cite{Echter2024} for an {\it equally} spaced SP spectrum.
As mentioned above, the latter has shown that the MBDOS as a function of the excitation energy $Q$ does not distinguish between fermionic and bosonic systems. 
Furthermore, as the number of particles $N$ increases for a fixed $Q$, a point is reached where all excitation quanta are distributed among all particles ($\rho_0 Q=N$). 
Beyond this point, adding new particles does not create new MB configurations. 
Instead, it merely increases the occupancy of the energetically lowest avaliable SP states (depending on the particles statistics). 
Consequently, the MBDOS of systems with rigid SP level spacing saturates and becomes independent of $N$ for a given $Q$. 
We now investigate whether these findings hold in the presence of SP spectral fluctuations.
Figure \ref{fig:eff_D} displays  $\langle\mathcal{N}^{(\pm)}(N,Q)\rangle$ and $\delta\mathcal{N}^{(\pm)}(N,Q)$ for $N_{\rm R} = 10^5$ realizations at $\rho_0Q=5$ as a function of $N$.

\subsubsection{{Fermionic case}}

For fermionic systems, both $\langle\mathcal{N}^{(-)}(N,Q)\rangle$ and $\delta\mathcal{N}^{(-)}(N,Q)$ saturate, regardless of whether the underlying SP dynamics is integrable or chaotic, as seen in panels (a) to (d).
This suggests that the saturation in fermionic systems is a general property of individual realizations within the ensemble, rather than a consequence of the ensemble average.

In the case of chaotic SP dynamics, the saturation value  of $\langle\mathcal{N}^{(-)}\rangle$ approaches that  of the Bethe approximation, with increasing rigidity of the universality class, as shown in panels (b) to (d).
Note that $\delta\mathcal{N}$ remains relatively small compared to the mean.
In contrast, for integrable SP dynamics, see panel (a), both the mean and the variance of $\mathcal{N}^{(-)}$ exhibit significantly larger values, in line with what is observed in Fig.~\ref{std_for_each_ensemble}. 

Based on our findings, we propose that for fermionic systems, the (smooth) MB counting function (or DOS) generally varies depending on whether the underlying SP system is integrable or chaotic. 
In the chaotic case, we anticipate 
close agreement with the Bethe approximation, whereas for integrable systems, such agreement is not necessarily expected.

\subsubsection{{Bosonic case}}

For bosonic systems with chaotic SP dynamics, shown in panels (f) to (h), the mean saturates, and both the saturation value and the corresponding number of particles increase as the SP level repulsion decreases. 
Notably, in the GOE case, see panel (f), $\delta\mathcal{N}^{(+)}(N,Q)$ has not yet saturated for $N<550$, although its growth appears to be slowing down. 
In the other chaotic cases, see panels (g) and (h), $\delta\mathcal{N}^{(+)}(N,Q)$ does saturate, and the corresponding number of particles needed for saturation increases as the SP level repulsion decreases.
Based on this trend, it is reasonable to expect that $\delta\mathcal{N}^{(+)}(N,Q)$ would also saturate for systems with GOE statistics, albeit at a larger particle number than $550$.

In contrast, for bosonic systems with integrable SP dynamics, neither $\langle\mathcal{N}^{(+)}(N,Q)\rangle$ nor $\delta\mathcal{N}^{(+)}(N,Q)$ saturate for up to $N \approx 400$ particles.
In \cite{Echter2024}, it was shown that the saturation of the MBDOS with increasing number of bosons can be attributed to the gap $\Delta_1$ between the first excited SP state and the SP ground state.
When $\Delta_1\neq0$, there is a threshold for the number of particles, $N_{\rm th}\geq Q/\Delta_1$, beyond which the counting function saturates.
This implies that for a SP system with an exactly degenerate ground state ($\Delta_1$=0), in the limit of $\rho_0 Q \to 0$, the number of MB states grows linearly with $N$ and no saturation of the counting function occurs. 
Since SP systems with exactly degenerate ground-state energies are unusual, we removed those from the ensemble used for panel (e), although this did not change the 

saturation behavior of $\langle\mathcal{N}\rangle$, nor $\delta\mathcal{N}$, for the parameters chosen in panel (e). 

This peculiar behavior in the integrable SP case can be attributed to the absence of spectral correlations between SP levels, leading to frequent 
quasi-degenerate SP states, with increasing probability for decreasing SP level spacing.
As a result, the ensemble will be dominated by realizations where $\Delta_1 \to 0$, for which $N_{\rm th}\to \infty$. Hence no saturation of $\langle\mathcal{N}^{(+)}(N,Q)\rangle$ and of $\delta\mathcal{N}^{(+)}$ is expected,  in line with the behavior of $\langle\mathcal{N}^{(+)}(N,Q)\rangle$ and $\delta\mathcal{N}^{(+)}$ in the Poisson case (panel (e) in Fig.~\ref{fig:eff_D}), and contrary to the case of a rigid spectrum \cite{Echter2024}.

Our findings imply that, for bosonic systems with chaotic SP dynamics, the saturation observed with increasing $N$ at fixed $Q$ is not limited to the ensemble average, but is a general property of each individual realization within the (random matrix) ensemble of systems.

To summarize, we claim that in the case of bosonic systems, the ensemble-averaged MB excitation energy counting function, respectively, MBDOS can differentiate between integrable and chaotic SP motion as well as the corresponding variances.

\section{Conclusions and Outlook}
\label{sec:conclusions}

In this paper, we highlighted the difference between a mean MBDOS obtained by averaging the spectrum of an individual MB system using a pure energy average, on the one hand, and a mean MBDOS after ensemble average at fixed energy $E$, on the other hand. In our work, we focused on the latter case and consider systems of identical particles, fermions and bosons, in a mean-field framework. We demonstrated how spectral features of chaos versus integrability on the SP level impact the ensemble-averaged MBDOS and its variance.
We analyzed these quantities for two different physical settings:
First, as a function of the total MB energy $E$ and second, as a function of the excitation energy, $Q=E-E_{\rm GS}$, where $E_{\rm GS}$ is the energy of the ground state, a setting which is closer to experimental information.
Eliminating in this way fluctuations in $E_{\rm GS}$ significantly affects the ensemble-averaged results.

In general, we observed that the ensemble-averaged MBDOS and its variance are significantly influenced by both the fermionic and bosonic character as well as the fluctuations of the underlying SP spectra, distinguishing between integrable and chaotic SP dynamics.

Starting with the ensemble average at fixed $E$ we first derived closed-form expressions for the average mean-field MBDOS of indistinguishable particles, formulated in terms of 
nested convolutions of 
cluster functions associated with
the corresponding SP spectral universality classes.
For uncorrelated (Poisson-type) SP level fluctuations, we were able to derive explicit analytical expressions for the MBDOS.
(We obtained additional insight from further evaluating these expressions to obtain closed analytical results for $N=2$ particle systems.)

Interestingly, we generally found that for fermions the SP level fluctuations precisely cancel all subleading contributions from indistinguishability. 
Hence, the average MB spectral density reduces to the (Thomas-Fermi) volume term, in contrast to the bosonic case.
The analytical expressions were  validated through numerical simulations.
In other cases, where analytical progress was not feasible, we relied on numerical simulations for insights.

Our analytical analysis, combined with our extensive numerical simulations for systems with several particles, consistently reveal significant differences, namely factors of 4 up to about 10 for fermions and even larger for bosons, in the MB-body level densities and their variances, depending on the nature of the underlying dynamics. 

Our numerical simulations to investigate the ensemble-averaged MBDOS, where each realization was evaluated at its excitation energy $Q$, lead to the following results: 
With increasing $Q$ the cumulative MBDOS, the level counting function, exhibits huge fluctuations for the Poisson case such that the mean counting function looses its meaning. However, SP level correlations significantly reduce the size of the MB fluctuations highlighting the role of the mean MBDOS and, for instance, explaining the success of the Bethe formula for the MBDOS: 
For fermionic systems exhibiting chaotic SP motion, fluctuations are suppressed and we found good agreement with the well-established Bethe formula. 

Furthermore, we studied the $N$-dependence of the ensemble-averaged cumulative MBDOS and its variance, for fixed excitation energy $Q$.
Analogous to findings in systems with equally spaced SP levels, we observed a saturation behavior that is sensitive to both the SP dynamics and the particle statistics. Our results imply that the ensemble-averaged MB excitation energy counting function as well as
the corresponding variances can differentiate between the underlying integrable and chaotic SP dynamics. The differences are drastic in particular for bosons where the average counting function and its variance saturate for random matrix-like SP dynamics, in contrast to the SP Poisson case.

A natural extension of our study would involve relaxing the constraint of a constant mean SPDOS and considering an energy-dependent mean level spacing instead. 
This would allow for the investigation of the ergodic hypothesis and the saturation of the MBDOS with increasing particle number, within a broader physical framework.
We note that while our report focuses on the non-interacting limit, we anticipate that extending this work to interacting systems will reveal a wealth of intriguing physics and is a research path that we plan to pursue. 

Regarding possible applications of our findings, 
considering that the smooth part of the SPDOS is a fundamental tool for analyzing SP dwell times in scattering systems \cite{Carvalho2002, Lewenkopf2004}, we envision that the ensemble-averaged MBDOS can serve a similar role in studying MB dwell times. 
Our ensemble-averaged MBDOS naturally emerges as a candidate to represent this smooth part, especially in non-interacting or weakly interacting systems.

Furthermore, in the introduction we mentioned a disorder average of free fermions in  closed mesoscopic systems as an application and prominent example of the ensemble average discussed in detail throughout this paper. Possible interesting implications of our results for disordered systems in the localized (Poisson) and delocalized (random matrix) regime are left for future work.

\begin{acknowledgments}
The authors thank Steven Tomsovic for useful discussions.
GM is funded through a fellowship by the Studienstiftung des Deutschen Volkes.
CL thanks the Brazilian funding agencies CNPq and FAPERJ for their support. 
We acknowledge further financial support from the Deutsche Forschungsgemeinschaft (German Research Foundation) through Ri681/15-1 (project number 456449460) within the Reinhart-Koselleck Programme.
\end{acknowledgments}

\appendix

\section{Integrable case}

Here we provide further details of the derivation of Eq.~\eqref{final_poisson} starting from Eq.~\eqref{poisson_prelimiary}. 
For clarity we consider only the integrals of Eq.~\eqref{poisson_prelimiary}, omitting the summation over the coefficients $\tilde{c}$. 
By explicitly expressing the corresponding expression in terms of $E_i = x_i N_i$, we obtain
\begin{widetext}
\begin{equation}\label{Poisson_integration}
    \begin{split}
        \int_0^\infty\!\!\text{d}E_1 \cdots\!\int_0^\infty\!\!&\text{d}E_l  \,\delta\!\left(E - \sum_{j=1}^l E_j\right) \prod_{i=1}^m\prod_{n=1}^{\left|L_i\right|-1}
        \delta\!\left(\frac{E_{k_{i,n}}}{N_{k_{i,n}}}-\frac{E_{k_{i,n+1}}}{N_{k_{i,n+1}}}\right) \\
        =&\left(\prod_{i=1}^m \prod_{n=1}^{\left|L_i\right|} N_{k_{i,n}}\right) \int_0^\infty\text{d}E_1 \cdots\int_0^\infty\text{d}E_l~
        \delta\!\left(E - \sum_{i=1}^m \sum_{n=1}^{\left|L_i\right|} N_{k_{i,n}}E_{k_{i,n}}\right) \prod_{i=1}^m \prod_{n=1}^{\left|L_i\right|-1}\delta\!\left(E_{k_{i,n}}-E_{k_{i,n+1}}\right) \\
        =&\left(\prod_{i=1}^m \prod_{n=1}^{\left|L_i\right|} N_{k_{i,n}}\right) \int_0^\infty\text{d}E_{j_1}\cdots \int_0^\infty\text{d}E_{j_m}~\delta\!\left(E - \sum_{i=1}^m E_{j_i}\sum_{n=1}^{\left|L_i\right|} N_{k_{i,n}}\right) \\
        =&\left(\prod_{i=1}^m \frac{\prod_{n=1}^{\left|L_i\right|} N_{k_{i,n}}}{\sum_{n=1}^{\left|L_i\right|} N_{k_{i,n}}}\right) \int_0^\infty\text{d}E_{j_1}\cdots \int_0^\infty\text{d}E_{j_m}~\delta\!\left(E - \sum_{i=1}^m E_{j_i}\right) =\left(\prod_{i=1}^m \frac{\prod_{n=1}^{\left|L_i\right|} N_{k_{i,n}}}{\sum_{n=1}^{\left|L_i\right|} N_{k_{i,n}}}\right) \frac{E^{m-1}}{(m-1)!}.
    \end{split}
\end{equation}
\end{widetext}
Using Eqs.~\eqref{permutations_in_cycle}, \eqref{tilde_c} and \eqref{poisson_coefs}, we can straightforwardly arrive at 
\begin{equation}
    \begin{split}\label{all_of_the_cs}
        \tilde{c}(N_1,\dots,N_l)\left(\prod_{i=1}^m \frac{\prod_{n=1}^{\left|L_i\right|} N_{k_{i,n}}}{\sum_{n=1}^{\left|L_i\right|} N_{k_{i,n}}}\right)&=\\
        \frac{1}{N!}\ c_{L_1,...,L_m}\!\left(N_1,...,N_l\right)&\ c(N_1,\dots,N_l).
    \end{split}
\end{equation}
The combination of Eq.~\eqref{Poisson_integration} with \eqref{all_of_the_cs} provides a path to obtain Eq.~\eqref{final_poisson} from Eq.~\eqref{poisson_prelimiary}.

\subsection{Average MBDOS of a boson gas with Poisson SP level-spacing statistics}
\label{app:poisson_bososns}

Equations~\eqref{permutations_in_cycle}, \eqref{final_poisson}, and \eqref{poisson_coefs} allow one to obtain 
$\langle\rho^{(+)}(N,E)\rangle_P$ in a closed-form analytical expression. 
Below, we present $\langle\rho^{(+)}(N,E)\rangle_P$ for several selected cases: 
\begin{equation}
\label{poissonN2358}
    \begin{split}
        \langle\rho^{(+)}(N\!=\!2,E)\rangle_P&\!=\frac{\rho_{0}^2E}{2} \!+\! \frac{1}{2}\rho_{0}\\
        \langle\rho^{(+)}(N\!=\!3,E)\rangle_P&\!=\frac{\rho_0^3 E^{2}}{12} \!+\! \frac{\rho_{0}^2E }{2} + \frac{1}{3}\rho_{0}\\
        \langle\rho^{(+)}(N\!=\!5,E)\rangle_P&\!= \frac{\rho_{0}^{5}E^{4} }{2880} \!+\! \frac{\rho_{0}^{4}E^{3} }{72} \!+\! \frac{7 \rho_{0}^{3} E^{2} }{48} \!+\! \frac{5 \rho_{0}^2 E }{12} \!+\!\frac{1}{5}\rho_{0}\\
        \langle\rho^{(+)}(N\!=\!8,E)\rangle_P&\!=\frac{\rho_{0}^{8} E^{7} }{203212800} \!+\! \frac{\rho_{0}^{7} E^{6} }{1036800} \!+\! \frac{23 \rho_{0}^{6} E^{5} }{345600} \\
        &
        \!\! \!\!\!\!\!\!\!\!\!\!\!\!\!\!\!\!\!\!\!\!\!\!\!\!\!\!\!\!\!\!\!
        +\frac{7 \rho_{0}^{5} E^{4} }{3456} + \frac{967  \rho_{0}^{4} E^{3}}{34560} + \frac{469 \rho_{0}^{3} E^{2} }{2880} + \frac{363 \rho_{0}^2 E }{1120} + \frac{1}{8}\rho_{0}.
    \end{split}
\end{equation}

While the MBDOS can be obtained analytically (with reasonable effort) for $N=2$ and $N=3$, the algebra becomes prohibitively extensive as $N$ increases. 
For this reason, we implemented a numerical algorithm to compute $\langle\rho^{(+)}(N,E)\rangle_P$ for $N>3$.

\subsection{Average MBDOS of a fermion gas with Poisson SP level-spacing statistics}
\label{app:poissom_fermions}

Here, we provide further details on the derivation of the ensemble-averaged MBDOS for fermionic systems with SP level spacing that obeys Poisson statistics $\langle \rho^{(-)}(N,E)\rangle_P$. 
As mentioned in the main text, only the leading order in the Weyl expansion survives. 
Rewriting Eq.~\eqref{final_poisson} after substituting Eqs.~\eqref{permutations_in_cycle} and \eqref{poisson_coefs}, we find that $\langle \rho^{(-)}(N,E)\rangle_P$ takes the following form
\begin{equation}
\begin{split}
    \left\langle\rho^{(-)}(N,  E)\right\rangle_P &= \frac{1}{N!} \sum_{m=1}^N \frac{E^{m-1}}{(m-1)!}\rho_0^m \\ &\!\!\!\!\!\!\!\!\!\!\!\!\sum_{\substack{n_1,...,n_m = 1 \\ n_1 \leq ... \leq n_m \\ \sum_{i=1}^m n_i = N}}^N \left(\prod_{i=1}^m \frac{1}{n_i}\right) d\!\left(N;n_1,...,n_m\right)
\end{split}
\end{equation}
with
\begin{equation}
\begin{split}
\label{app:modified_stirling_orthogonality}
&d\!\left(N;n_1,...,n_m\right) = \sum_{l=m}^N (-1)^{N-l} \\
&\sum_{\substack{N_1,...,N_l = 1 \\ N_1 \leq ... \leq N_l \\ \sum_{i=1}^l N_i = N}}^N \frac{N!}{\prod\limits_{N_i \in \left\{N_1,...,N_l\right\}} \!\!\!\! m\!\left(N_i\right)!  \prod\limits_{i=1}^l N_i} \sum_{\substack{L_1 \sqcup ... \sqcup L_m = \{1,...,l\} \\ \sum_{k \in L_i} N_k = n_i}} \! 1.
\end{split}
\end{equation}
We define the modified Stirling numbers
\begin{equation}
\label{app:s}
(-1)^{N-l} s^N_{\left[N_1,...,N_l\right]} = \frac{N!}{\prod\limits_{N_i \in \left\{N_1,...,N_l\right\}} \!\!\!\! m\!\left(N_i\right)! \left(\prod\limits_{i=1}^l N_i \right)}
\end{equation}
being the number of permutations of $N$ entities with cycle decomposition into $l$ cycles of lengths $N_1,...,N_l$ and
\begin{equation}
\label{app:S}
S^{\left[N_1,...,N_l\right]}_{\left[n_1,...,n_m\right]} = \sum_{\substack{L_1 \sqcup ... \sqcup L_m = \{1,...,l\} \\ \sum_{k \in L_i} N_k = n_i}} 1
\end{equation}
the number of set partitions of a set with $l$ elements into $m$ non-empty disjoint subsets under the constraint that $\sum_{k \in L_i} N_k = n_i$ for all $1 \leq i \leq m$ after possibly 
renumbering the $L_i$
\footnote{ We require the existence of a one-to-one assignment $j: \{1,...,m\} \to \{1,...,m\}$ such that $\sum_{k \in L_{j(i)}} N_k = n_i$ holds for all $i \in \{1,...,m\}$. Then by reindexing the $L_i$, concretely by applying the bijective map $j^{-1}$ to the index set, we can achieve that $\sum_{k \in L_i} N_k = n_i$ for all $i \in \{1,...,m\}$. We count only up to reindexing.}
.

Combining Eqs.~\eqref{app:modified_stirling_orthogonality},\eqref{app:s} and \eqref{app:S} yields
\begin{equation}
d\!\left(N;n_1,...,n_m\right) = \sum_{l=m}^N \sum_{\substack{N_1,...,N_l = 1 \\ N_1 \leq ... \leq N_l \\ \sum_{i=1}^l N_i = N}}^N s^N_{\left[N_1,...,N_l\right]}S^{\left[N_1,...,N_l\right]}_{\left[n_1,...,n_m\right]}.
\end{equation} 
Further, we postulate their orthogonality 
\begin{equation}
\label{orthogonality}
d\!\left(N;n_1,...,n_m\right) = \begin{cases} 1 & m = N \\ 0 & m \neq N \end{cases}
\end{equation}
when $\sum_{i=1}^m n_i = N$.
We have verified the validity of Eq.~\eqref{orthogonality} up to $N=10$ by numerically evaluating Eq.~\eqref{app:modified_stirling_orthogonality}.

To the best of our knowledge, these objects have not been addressed in the
literature so far. We name them {\it modified Stirling numbers} due to their similarity to the signed Stirling numbers, which obey a similar orthogonality relation.

\section{First Level Statistics}
\label{app:first_level}
Several measures of spectral statistical properties focus on the energy levels contained within a given energy window. 
To obtain a suitable spectral interval $[\epsilon_-,\epsilon_+]$, the system's lowest energy level $\epsilon_0$ within this interval must be uncorrelated with $\epsilon_-$. 
This is achieved by introducing $\epsilon_-$ at a random energy within the spectrum. 
The difference to the subsequent level is regarded as the first level spacing while the following remain unchanged. 
The probability of placing $\epsilon_-$ between the $n$-th and $n+1$-th level is proportional to $l_n/(N\langle l \rangle)$, with $l_n = \epsilon_{n+1} - \epsilon_n$,
$\langle l\rangle = \frac{1}{N} \sum_n l_n$ is the spectral mean level spacing, and $N$ the total number of levels. 
Assuming the probability of a level spacing $l_n$ is $\bar{p}(l_n)$, the expectation value of the spacing in which $\epsilon_-$ is placed is given by 
\begin{equation}
    \sum_{n=1}^N \int_0^{N\langle l\rangle} \text{d}l_n~ l_n \frac{l_n}{N\langle l\rangle} \bar{p} (l_n) \xrightarrow[N\to \infty]{} \frac{1}{\langle l\rangle} \int \text{d}l ~l^2 \bar{p} (l).
\end{equation}
In the limit of $N\to\infty$ it is straightforward to see that the conditional probability of finding a level spacing of length $l$ in which $\epsilon_-$ is placed is given by 
\begin{equation}
    p(l)=\frac{l}{\langle l\rangle} \bar{p}(l).
\end{equation}
Let us now consider an interval of length $l$ that contains $\epsilon_-$. 
The probability of placing $\epsilon_-$ within this interval such that the difference to the new first level is $\epsilon_0$ is given by $\Theta(l-\epsilon_0)/l$.
Finally, the probability of finding a first level spacing, namely $\epsilon_0-\epsilon_-$, is
\begin{equation}
        P(\epsilon_0)=\int_0^\infty \!\!\text{d}l~ \frac{\Theta(l-\epsilon_0)}{l} \frac{l}{\langle l \rangle} \bar{p}(l)
        =\frac{1}{\langle l\rangle} \int_0^\infty \!\! \text{d}a~\bar{p}(a+\epsilon_0),
\end{equation}
where $l=a+\epsilon_0$.

\section{Cluster functions}
\label{app:cluster_fktS}

The cluster functions $Y_n(r)$ used in Sec.~\ref{sec:Chaos} are taken from Ref.~\cite{bohigas1991random}, 
and are valid for all values of $r$.
This is crucial because Eq.~\eqref{general_final} involves integrals of $Y_n(r)$ over the entire domain of $r$.
We emphasize that the standard RMT reference \cite{mehta2004random} provides cluster functions that are only valid for $r\geq0$.
Neglecting this restriction and using $Y_n(r)$ as provided by Mehta \cite{mehta2004random} leads to incorrect results for the GOE case, similar to those reported in Ref.~\cite{Munoz2006} 
(which also overlooked the contact terms discussed in Sec.~\ref{sec:results_integrable}).

Fortuitously, this limitation of Mehta's $Y_n(r)$ does not affect the GUE and GSE cases for $N=2$.
This can be observed through a direct inspection of the two-point cluster functions $Y_2(r)$, which are given by \cite{bohigas1991random} 
\begin{eqnarray}
    Y_2^\text{GOE}(r)&=&\left[\frac{\sin(\pi r)}{\pi r}\right]^2 \nonumber \\
    &-& \left[\text{Si}(\pi r)-\pi \epsilon(r)\right] \left[\frac{\cos(\pi r)}{\pi r}-\frac{\sin (\pi r)}{(\pi r)^2}\right], \nonumber\\
    Y_2^\text{GUE}(r)&=&\left[\frac{\sin(\pi r)}{\pi r}\right]^2, \\
    Y_2^\text{GSE}(r)&=&\left[\!\frac{\sin(\pi r)}{\pi r}\!\right]^2\!\!- {\rm Si}(2\pi r)\!\left[\!\frac{\cos(2\pi r)}{2\pi r} -\frac{\sin(2\pi r)}{(2\pi r)^2}\!\right], \nonumber
\end{eqnarray}
where $\text{Si}(r)$ is the sine integral and
\begin{equation}
    \epsilon(r)=
    \begin{cases}
        \;\,\,\frac{1}{2} & r>0\\
        \;\;\,0 & r=0\\
        -\frac{1}{2} & r<0.
    \end{cases}
\end{equation}

\begin{figure}
    \centering
    \includegraphics[width=\linewidth]{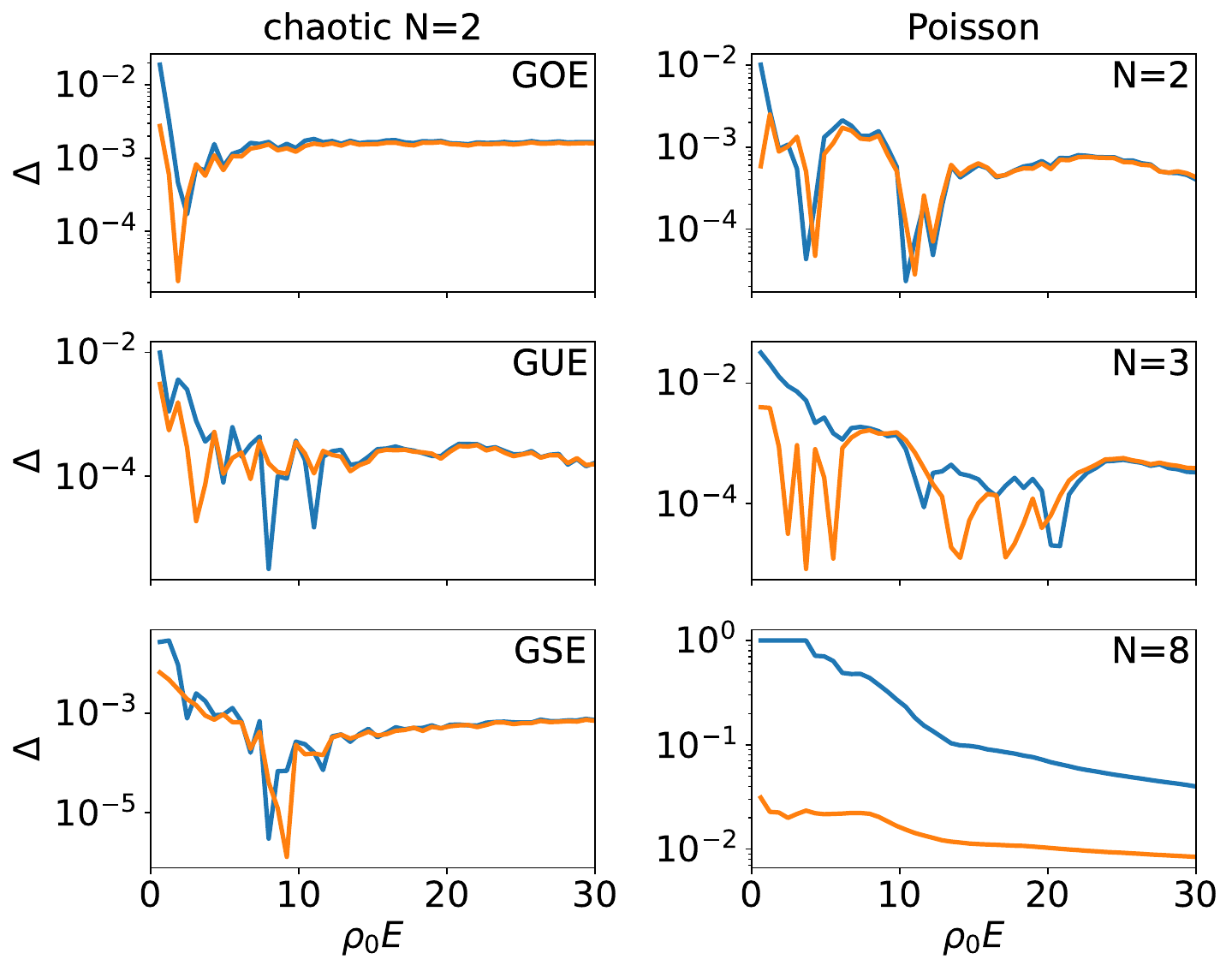}
    \caption{
    Absolute relative deviation $\Delta$ between the numerical simulation, with $N_{\rm R}=10^5$, and the analytically predicted $\langle {\cal N}^{(\pm)} (E, N)\rangle$ for fermions (blue) and bosons (orange) as a function of the energy $E$. 
    Left column: $N=2$ particle systems with underlying chaotic dynamics for the random matrix universality classes $\beta=1,2,4$. 
    Right column: Systems with underlying integrable SP dynamics with $N=2,3\text{ and }8$ particles.
    }
    \label{relativ_error}
\end{figure}

\section{Comparison between numerical simulations and analytical results}
\label{app:error-analysis}

In this Appendix we analyse the precision of numerical simulations to compute
$\langle {\cal N}^{(\pm)} (E, N)\rangle$.

Figure \ref{relativ_error} depicts the relative deviation $\Delta$ of the average integrated MBDOS, $\langle {\cal N}^{(\pm)} (E, N)\rangle$, between numerical simulations and analytical results, 
given by Eqs. \eqref{poissonN2358}, \eqref{WV_poisson}, and \eqref{DOS_RMT_N2},
for chaotic and integrable SP motion. 
As expected the numerical average displays larger relative deviations for small energies, where the level density is low. 
This effect is more pronounced in fermionic systems since part of the energy corresponds to the formation of a MB ground state, $E_{GS}$. 
Notably, since the fermionic $E_{GS}$ increases with increasing particle number, the relative deviation grows accordingly. 
In the case of $N=8$ fermions with integrable SP motion, the increase in MB ground state energy is so significant that we did not find a single realization in the numerical simulation where $E_{\rm GS}\lesssim\rho_05$, resulting in a relative deviation of order 1.
The small values obtained for $\Delta$ support the use of numerical simulations as an effective tool to analyze the ensemble-averaged MB counting function and DOS for systems with underlying chaotic SP dynamics for $N>2$.


\bibliography{MB-DOS}

\end{document}